\documentclass[twocolumn]{aastex6}
\usepackage[]{units}
\usepackage[update,prepend]{epstopdf}
\usepackage{graphicx}
\usepackage{amssymb}
\usepackage{amsmath}
\usepackage{threeparttable}
\usepackage{color}

\usepackage{mathtools}

\newcommand{\wise}{{\it WISE}}

\newcommand{\oiii}{\hbox{[O\,{\scriptsize III}]}}

\newcommand{\CIV}{C\,{\scriptsize IV}}
\newcommand{\NV}{N\,{\scriptsize V}}
\newcommand{\Lya}{Ly\,$\alpha$}

\newcommand{\uJy}{\,$\mu$Jy}
\newcommand{\um}{\,$\mu$m}

\newcommand{\kms} {\,km\,s$^{-1}$}
\newcommand{\ergs} {\,erg\,s$^{-1}$}
\newcommand{\Lsun}{\,L{\scriptsize $\odot$}}
\newcommand{\Msun}{\,M$_\odot$}
\newcommand{\Msunyr}{\,M$_\odot$\,yr$^{-1}$}
\newcommand{\ang}{\,\AA}
\newcommand{\pbeam}{\,beam$^{-1}$}

\mathchardef\mhyphen="2D
\shorttitle{Winds as the origin of radio emission}
\shortauthors{Hwang et al.}

\graphicspath {{figs/}, {figs/0510_LoBAL/}}

\begin{document}

\title{Winds as the origin of radio emission in $z=2.5$ radio-quiet extremely red quasars}

\author{Hsiang-Chih Hwang\altaffilmark{1}, Nadia L. Zakamska \altaffilmark{1}, Rachael M. Alexandroff\altaffilmark{2,3}, Fred Hamann\altaffilmark{4,5}, Jenny E. Greene\altaffilmark{6}, Serena Perrotta\altaffilmark{4}, Gordon T. Richards\altaffilmark{7}}

\altaffiltext{1}{Department of Physics \& Astronomy, Johns Hopkins University, Baltimore, MD 21218, USA}
\altaffiltext{2}{Canadian Institute for Theoretical Astrophysics, The University of Toronto, Toronto, ON M5S 3H8, Canada}
\altaffiltext{3}{Dunlap Institute for Astronomy and Astrophysics, The University of Toronto, Toronto, ON M5S 3H4, Canada}
\altaffiltext{4}{Department of Physics and Astronomy, University of California, Riverside, CA 92507, USA}
\altaffiltext{5}{Department of Astronomy, University of Florida, Gainesville, FL 32611, USA}
\altaffiltext{6}{Department of Astrophysical Sciences, Princeton University, Princeton, NJ 08544, USA}
\altaffiltext{7}{Department of Physics, Drexel University, Philadelphia, PA 19104, USA}

\begin{abstract}
Most active galactic nuclei (AGNs) are radio-quiet, and the origin of their radio emission is not well-understood. One hypothesis is that this radio emission is a by-product of quasar-driven winds. In this paper, we present the radio properties of 108 extremely red quasars (ERQs) at $z=2-4$. ERQs are among the most luminous quasars ($L_{bol} \sim 10^{47-48}$\ergs) in the Universe, with signatures of extreme ($\gg 1000$\kms) outflows in their \oiii$\lambda$5007\ang\ emission, making them the best subjects to seek the connection between radio and outflow activity. All ERQs but one are unresolved in the radio on $\sim 10$\,kpc scales, and the median radio luminosity of ERQs is $\nu L_\nu [{\rm 6\,GHz}] = 10^{41.0}$\ergs, in the radio-quiet regime, but one to two orders of magnitude higher than that of other quasar samples. The radio spectra are steep, with a mean spectral index $\langle \alpha \rangle = -1.0$. In addition, ERQs neatly follow the extrapolation of the low-redshift correlation between radio luminosity and the velocity dispersion of \oiii-emitting ionized gas. Uncollimated winds, with a power of one per cent of the bolometric luminosity, can account for all these observations. Such winds would interact with and shock the gas around the quasar and in the host galaxy, resulting in acceleration of relativistic particles and the consequent synchrotron emission observed in the radio. Our observations support the picture in which ERQs are signposts of extremely powerful episodes of quasar feedback, and quasar-driven winds as a contributor of the radio emission in the intermediate regime of radio luminosity $\nu L_\nu = 10^{39}-10^{42}$\ergs.
\end{abstract}

\keywords{galaxies: active -- quasars: general -- radio continuum: galaxies}

\section{Introduction}
\label{sec:intro}

Since the discovery of the first radio quasar 3C 273, the shape of the radio luminosity function of active galactic nuclei (AGNs) has remained a topic of active research. Measurements of the radio luminosity function of AGNs are critically important for understanding which physical processes contribute to the radio emission. While several authors suggest that the radio luminosity function of AGN is bimodal \citep{Kellermann1989, Xu1999, Ivezic2002}, or at least cannot be fit by a unimodal distribution \citep{Kratzer2015}, others find that the distribution of radio loudness indicators is instead continuous \citep{White2000, Rafter2009, LaFranca2010}. Determining the shape of the luminosity function is important for discerning whether multiple mechanisms may be responsible for the radio emission in different luminosity regimes. At a given AGN bolometric luminosity the radio power can span over six orders of magnitude -- thus mapping the radio luminosity function of various AGN subpopulations remains difficult and requires deep radio observations to probe the faint end -- the radio-quiet population.

While powerful jets dominate the bright radio sky \citep{Urry1995}, the radio-quiet population is still not well-understood. It is unclear whether star formation dominates the radio emission of radio-quiet AGNs \citep{DeVries2007, Kimball2011, Condon2013, Kellermann2016} or whether it is insufficient and other mechanisms are required \citep{Harrison2014, Rawlings2015, Zakamska2016b, Wang2017, White2017}. It may be a function of the power of the active nucleus, with star formation dominating the radio emission in low-power radio-quiet AGN \citep{Rosario2013}, but not in high-luminosity quasars \citep{Zakamska2016b}. 

Other possible contributors to the radio-quiet population are compact or low-power jets launched by central supermassive black holes \citep{Falcke2004, Leipski2006}, secondary emission from quasar-driven winds \citep{Stocke1992, Wang2008, Jiang2010, Ishibashi2011, FaucherGiguere2012, Zubovas2012, Zakamska2014, Nims2015}, or coronal emission \citep{Laor2008, Raginski2016}. The recently uncovered correlation between the velocity width of forbidden emission lines and the radio power within the radio-quiet population suggests a connection between radio emission and quasar-driven winds on large scales \citep{Mullaney2013, Zakamska2014}, making the secondary emission from quasar winds a particularly attractive possibility. In this picture, a wind produced on circumnuclear scales propagates into the surrounding interstellar medium and forms shocks which in turn generate relativistic electrons and the resulting synchrotron radiation. 

The recently discovered population of extremely red quasars (ERQs) provides an excellent opportunity to test whether radio emission may be associated with quasar-driven winds. ERQs are selected by their high mid-infrared-to-optical ratios \citep{Ross2015, Hamann2017}. With $L_{\rm bol} \sim 10^{47-48}$\ergs, ERQs are among the most luminous quasars at the peak epoch of quasar and star formation activity at $z=2-3$. Follow-up near-infrared spectroscopy of this population has revealed extremely broad and blueshifted \oiii$\lambda$5007\ang\ (hereafter \oiii) emission, indicating physical velocities of over 3000\kms\  (\citealt{Zakamska2016}; Perrotta et al. in prep). If these winds turn out to be spatially resolved on scales of a few\,kpc, they may represent the extreme end of galaxy-wide outflows detected in other quasars \citep{Liu2013, Brusa2015, Perna2015, Carniani2015}. 

In this paper, we present the radio properties of 108 ERQs identified by \cite{Hamann2017}. In Section \ref{sec:data} we describe our sample selection, observations, and data reduction. In Section \ref{sec:prop} we present the measurements of radio flux densities, spectral indices, and radio luminosities of ERQs. In Section \ref{sec:disc} we discuss the origin of radio emission in ERQs, and we conclude in Section \ref{sec:conc}. We use an $h=0.7$, $\Omega_m = 0.3$, $\Omega_\Lambda = 0.7$ cosmology. We use SDSSJhhmm+ddmm notation for our ERQ targets throughout the text and the full Jhhmmss.ss+ddmmss.s notation in Table \ref{table:flux}.

\section{Observations and data reduction}
\label{sec:data}

\subsection{Sample selection and observations}

Extremely red quasars (ERQs; \citealt{Hamann2017}) at redshifts $z=2.0 - 3.4$ are selected based on their high mid-infrared-to-optical ratios from optical and mid-infrared survey data. These sources are drawn from the Data Release 12 quasar catalog (DR12Q; \citealt{Paris2014, Paris2017}) of the Baryon Oscillation Spectroscopic Survey (BOSS; \citealt{Dawson2013}) which is part of the Sloan Digital Sky Survey - III (SDSS; \citealt{Eisenstein2011}), while the infrared fluxes are from the all-sky data of the {\it Wide-field Infrared Survey Explorer} (\wise; \citealt{Wright2010}).

The main sample of 97 ERQs is selected using two criteria, and these ERQs constitute $\sim1.6$ per cent of the number of luminous, blue quasars with similar redshifts and $W3$ magnitudes \citep{Hamann2017}. The first criterion is the red optical-to-mid-infrared color, specifically ${i_{\rm AB}} -{ W3_{\rm AB}} > 4.6$\,mag on the AB system where the $i$-band magnitudes are point-spread function magnitudes from SDSS Data Release 12 \citep{Alam2015} and $W3$ is the 12\um\ magnitude from \wise, with a conversion between AB and Vega system of $ W3_{\rm AB} = W3_{\rm Vega} + 5.242$ \citep{Cutri2011}. The color cut alone selects a variety of interesting sources, including heavily reddened quasars, low-redshift type 2 quasars, and broad absorption line quasars \citep{Ross2015}. In addition, \citet{Hamann2017} also require that the rest equivalent width of  \CIV $\lambda$1548,1551\ang\ (hereafter \CIV) be greater than 100\ang\ to preferentially select objects with suppressed continuum but strong emission lines, instead of normal quasars behind a dust reddening screen. To ensure robust measurements of emission line properties, sources with strong broad absorption features in \CIV\ are excluded. The selection requirements do not change as a function of redshift, and as a result slightly different populations may be selected at different redshifts. In practice, 75 per cent of the sample is confined to a relatively narrow range in redshift from 2.2 to 2.6. ERQs in this paper are referred to `core' ERQs in \cite{Hamann2017}, but here we remove `core' to avoid the confusion with the core in radio astronomy.

ERQs show peculiar properties including flat ultraviolet continua, wingless line shapes of \CIV, and unusual line flux ratios such as \NV $\lambda$1240\ang\ / \Lya\ $>$ 1 and high \NV / \CIV. Expanding on these interesting features of the ERQ population, \cite{Hamann2017} investigate an additional 235 ERQ-like objects. For this sample, instead of using a color cut and a \CIV\ rest equivalent width cut, \citet{Hamann2017} select objects with wingless \CIV\ profiles and find that this population generally favors red $i - W3$ colors and high \CIV\ equivalent widths, making them similar to ERQs. \citet{Hamann2017} suggest that the red colors of ERQs and their peculiar line properties might be related to powerful outflow activity and high accretion rates during a particular early phase of quasar evolution. 

In this paper, we investigate the radio properties of ERQs. Of the 97 ERQs, nine are detected at $F_{\nu}>1$\,mJy in the Faint Images of the Radio Sky at Twenty Centimeters (FIRST; \citealt{Becker1995}) survey, and their flux densities are listed in Table~\ref{table:FIRST-data}. We followed up the rest of the sample with sensitive radio observations using the Karl G. Jansky Very Large Array (VLA), but because the ERQ sample was still being finalized at the time, we targeted most ERQs and a few ERQ-like sources. Specifically, of the 88 ERQs without FIRST detections, 87 were observed while SDSSJ1212+5958 was not. In addition, we observed 14 ERQ-like objects, and one of them (SDSSJ1550+0806) is also detected in FIRST. SDSSJ1013+6112 matches all criteria of ERQs, but is listed in \citet{Hamann2017} as ERQ-like, and we keep this designation here.

\begin{figure}
\centering
\includegraphics[scale=0.43]{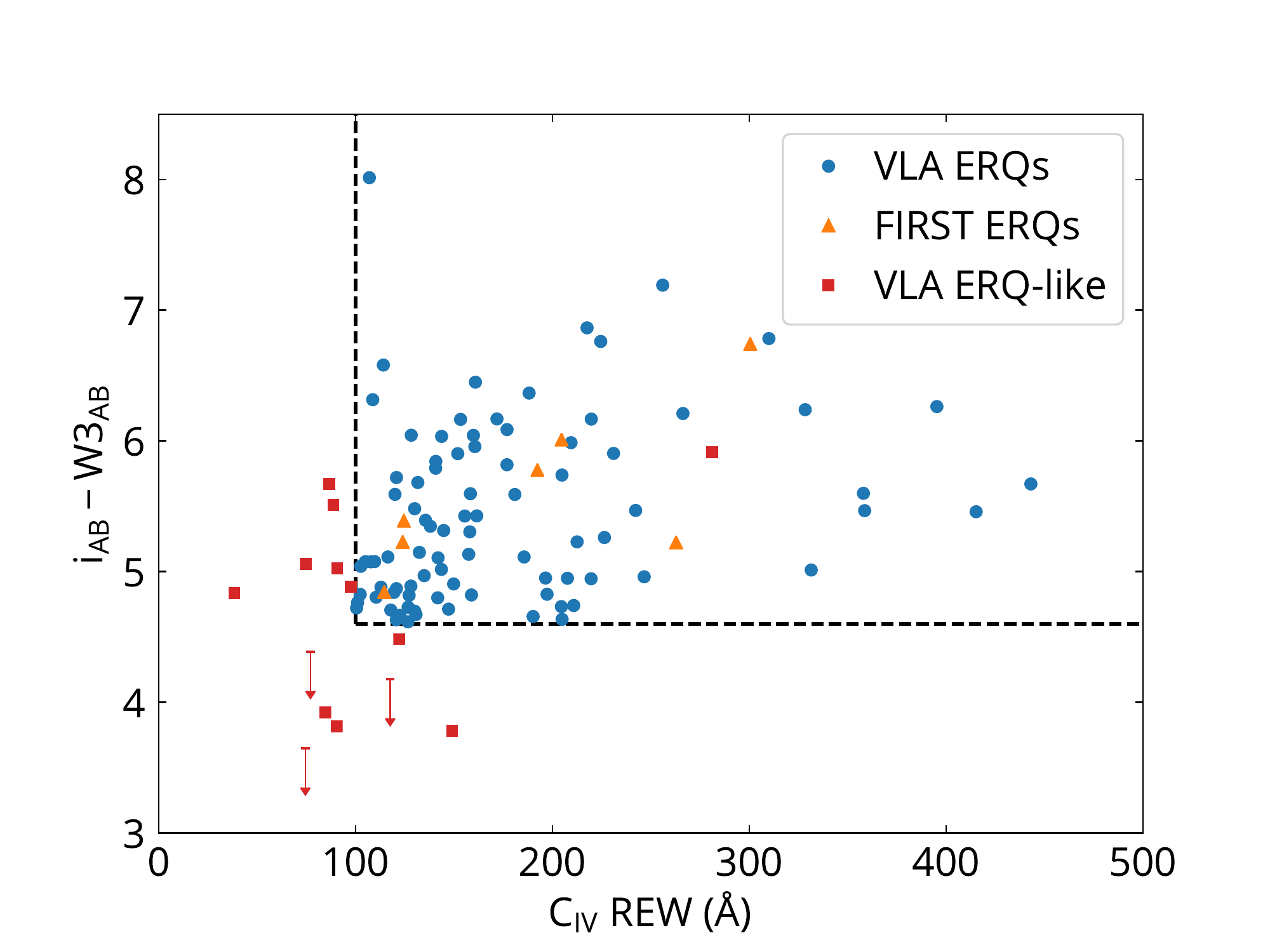} 
\includegraphics[scale=0.43]{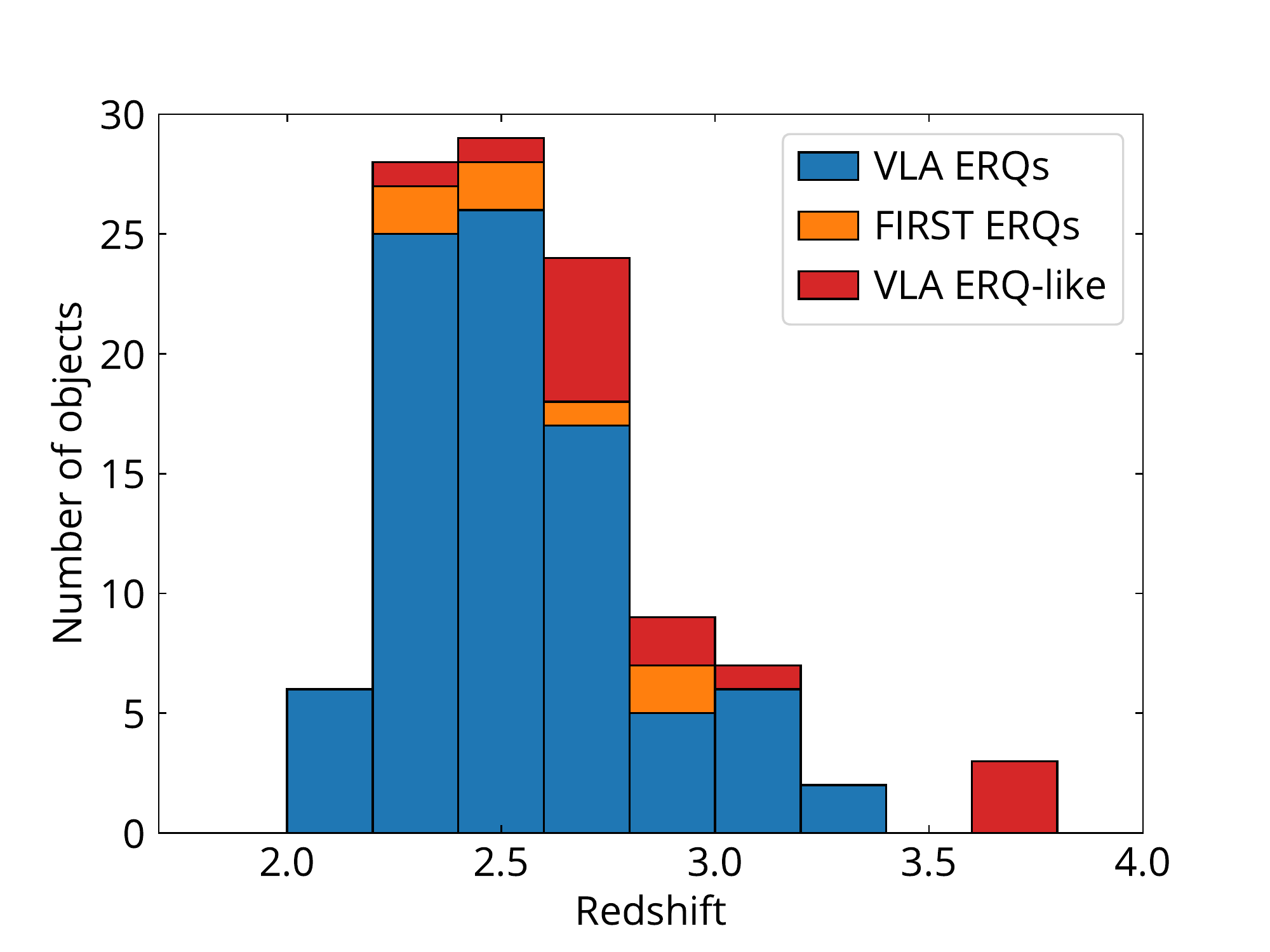} 
\caption{Colors and \CIV\ rest equivalent widths (REWs) (top) and redshifts (bottom) for the final 108 ERQ sample discussed in this paper, including 94 ERQs and 14 ERQ-like objects. AB magnitude is used for $i$ and $W3$. Blue dots are ERQs observed with VLA, orange triangles are ERQs detected in FIRST, and red squares represent ERQ-like objects. Three ERQ-like objects  (SDSSJ0844+5452, SDSSJ1447+5652, and SDSSJ1629+4957) are not detected in $W3$, and their upper limits of $W3$ are set to $W3_{\rm AB} = 17.3$, which is approximately the detection limit of $W3$ with a signal-to-noise ratio of 3. The dashed lines represent the selection criteria for ERQs, $i_{\rm AB} - W3_{\rm AB} > 4.6$ and rest equivalent width of \CIV\ larger than 100\ang.}
\label{fig:sample-selection}
\end{figure}

Two FIRST-detected ERQs, SDSSJ0915+5613 and SDSSJ0950+2117, are targeted for follow-up spectroscopy with SDSS only because of their radio emission detected by FIRST. Including such objects in calculations of the radio luminosity function would bias our results, so we exclude these two sources from all statistical analysis, though they are listed in Table~\ref{table:FIRST-data}. Thus, our final sample consists of 108 objects with a median redshift of 2.53 -- 7 FIRST-detected ERQs and 101 objects with targeted VLA observations of which 87 are ERQs and 14 are ERQ-like objects. The distribution in the color versus \CIV\ equivalent width space of the final sample is shown in Figure~\ref{fig:sample-selection}. As is apparent from the Figure, although ERQ-like objects are selected based on \CIV\ line shape, they also show the red colors and the high rest equivalent widths characteristic of the main ERQ sample, suggesting a strong connection between all these properties.

The observations were carried out using the VLA B configuration in the C-band ($\approx$ 4--8\,GHz; Program 16A-108, PI: Zakamska), which provides a resolution of 1.3 arcsec, or $\sim$ 10\,kpc at $z = 2.5$. The full bandwidth is 4\,GHz, with two basebands centered at 5.25 and 7.2\,GHz respectively to avoid radio frequency interference between 4 and 4.2\,GHz. All 101 objects were observed between June and August 2016 with 12 scheduling blocks. In each scheduling block, we started with a flux/bandpass calibrator, and after that gain calibrators and targets were observed alternately, with on-source time about 10 minutes for each target.

\subsection{Data reduction}
\label{sec:data-reduction}

\begin{figure*}
\centering
\includegraphics[height=1.9in]{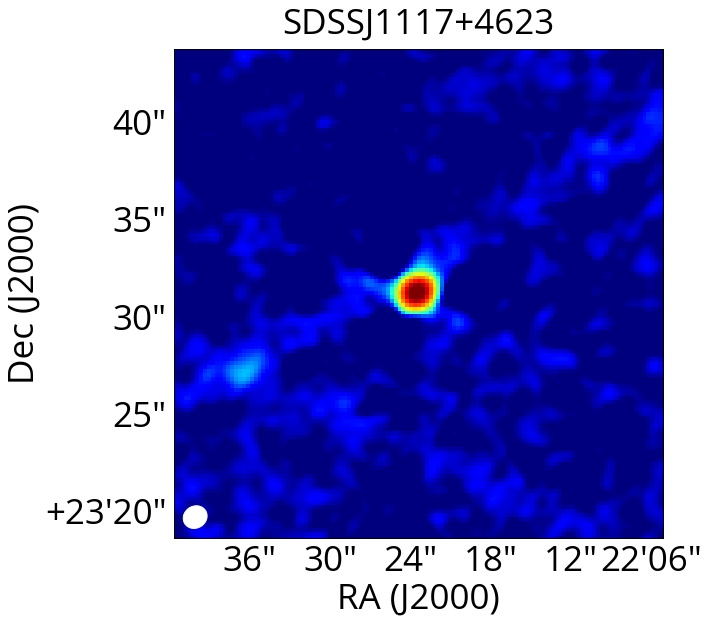}
\includegraphics[height=1.9in]{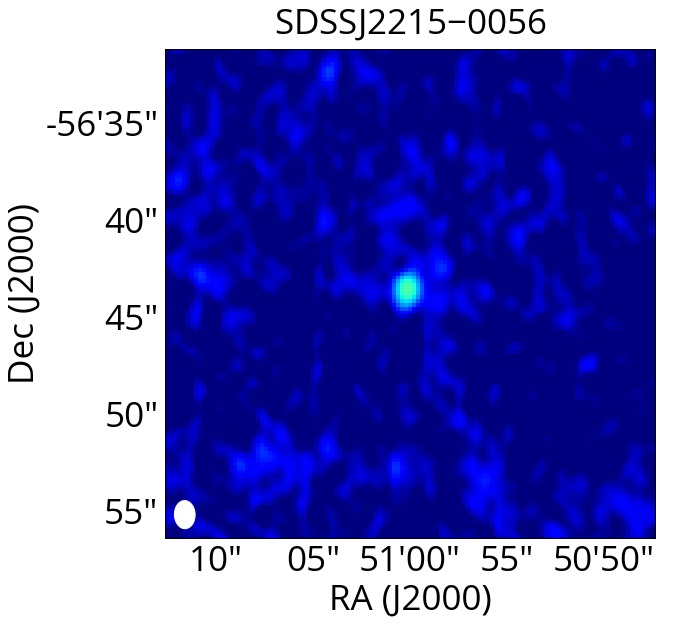} 
\includegraphics[height=1.9in]{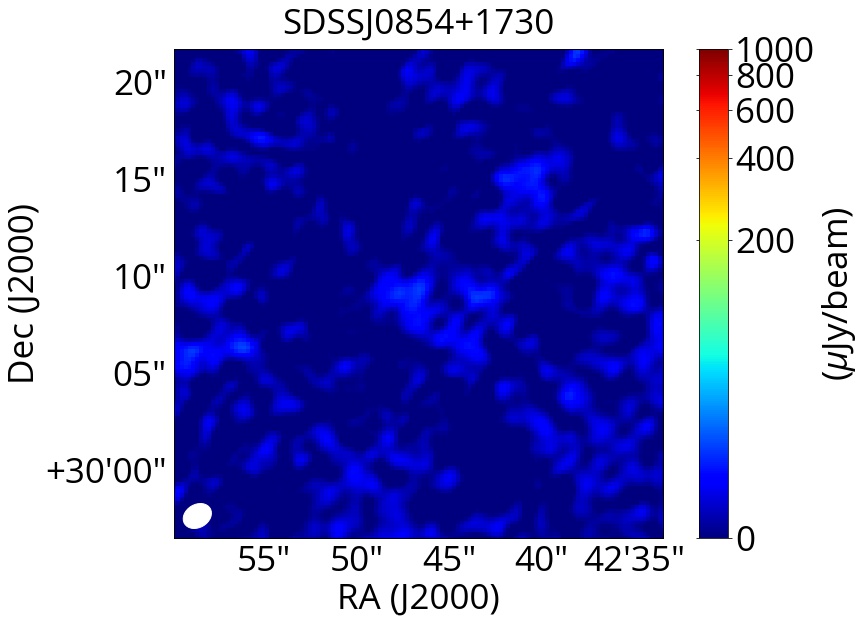} 
\caption{Three examples of cleaned images observed at 6.2\,GHz with the full-width-at-half-maximum (FWHM) ellipse of the restoring beam shown in the lower left corner. Left: SDSSJ1117+4623, the brightest object with a flux density of 1.3\,mJy in our VLA observations, with a beam of $1.27$ arcsec $\times1.13$ arcsec. A radio detection to the south-east does not have a counterpart in SDSS. It might be a $\sim 80$\,kpc jet lobe of SDSSJ1117+4623, or an unrelated source. Middle: SDSSJ2215$-$0056, an ERQ with a flux density of 98\uJy, typical of the ERQ sample, also detected in Stripe 82 at 1.4\,GHz. The size of the beam is $1.45$ arcsec $\times 1.06$ arcsec. As most of ERQs, the source is unresolved. Right: SDSSJ0854+1730, one of the non-detected objects with a beam of $1.53$ arcsec $\times 1.18$ arcsec. These three images share the same color scale, and therefore SDSSJ1117+4623 appears larger in the image due to extended wings of the Gaussian restoring beam, not because it is resolved. 
}
\label{fig:image-example}
\end{figure*}

We reduce the radio data from the VLA with Common Astronomy Software Applications (CASA) package v.4.7.0 \citep{McMullin2007}. Raw visibilities are flagged and calibrated by the pipeline, and all calibration solutions are inspected and manually flagged if needed. All images are cleaned with multi-frequency synthesis (MFS) mode with \texttt{nterms = 2}, which uses two terms in the Taylor expansion to model the sky spectrum and can reduce the artifacts due to wide-band imaging \citep{Rau2011}. We automatically make circular cleaning masks with a radius of 4 arcsec for those sources detected in the FIRST catalog \citep{Becker1995, White1997}. After the first round of cleaning, we inspect the cleaned maps and add additional masks for objects below the detection limit of the FIRST catalog and adjust the size of the masks manually if necessary. The final image product is obtained by repeated cleaning and modifying masks several times (usually 3 to 5 times) until the regions within $\sim1$ arcmin from our sources are free from dirty beams. In the meantime, we measure the root-mean-square noise level $\sigma$ of the cleaned images, and the final images are cleaned down to $4\sigma$, which is $\sim30$\uJy\pbeam\ with a median noise level of $7.61$\uJy\pbeam. 

For each radio image, CASA provides a theoretically derived point spread function (PSF). During these steps, we confirm that all detected sources are unresolved at the resolution of $\sim 1.3$ arcsec, except SDSSJ1117+4623 discussed in Section \ref{sec:indices}. Given that all sources are located at the image center with negligible angular size compared to the primary beam, we do not perform primary beam correction for MFS mode. Example cleaned images are shown in Figure~\ref{fig:image-example}.

Because all sources (except one) are unresolved, we use natural weighting to maximize the sensitivity of cleaned maps to point sources. However, if there is a source brighter than $\sim 100$\,mJy in the field (within $\sim$ 8 arcmin), or brighter than 10\,mJy and close to (within $\sim$1 arcmin) the target source, the limit of dynamic range as well as the stronger sidelobes due to natural weighting prevent us from measuring the flux densities robustly. To mitigate this problem, four objects are imaged with Briggs weighting (\texttt{ROBUST = 0.5} for SDSSJ0804+4701 and SDSSJ0935$-$0241, and \texttt{ROBUST = 0.0} for SDSSJ0936+1019 and SDSSJ2203+1219), and additional phase and gain self-calibrations are applied to reach a higher dynamic range. Using Briggs weighting lowers the sensitivity to a point source, so we avoid its use except in these four cases where it is warranted by the contaminating artifacts.

\subsection{Flux density measurement and CLEAN bias} 
\label{sec:flux-measurement}
To test the robustness of the flux density measurement, we insert an artificial point source in the central region of the field of SDSSJ1047+4844 with noise level $\sigma\sim7$\uJy\pbeam, where no source is detected within $\sim2$ arcmin of the field center. For a specific flux density, we randomly place a point source at a distance less than 10 arcsec from the center, and measure the flux density using several methods after cleaning down to $4\sigma$ level with MFS mode of \texttt{nterms = 2} and natural weighting, as described in Section~\ref{sec:data-reduction}. This routine is repeated 20 times for each flux density bin.

We present the results in Table~\ref{table:cleanbias}, where we record five flux density measurements for the artificial sources. In the first method, we report the maximum pixel intensity of the source in units of\uJy\pbeam. In the other four methods, the source is fit with an elliptical two-dimensional Gaussian and we record its peak and integrated flux density (in \uJy\pbeam and \uJy, respectively) with and without an additional parameter for the background sky level. If the image is free from noise and sidelobes, fitted peak and integrated flux densities of a point source should all be the same, while the results of the first method may be different since it depends on the sampling position of sources. 

We find that integrated flux densities usually overestimate the flux densities for a point source, which may be due to the uncleaned PSF under $4\sigma$ level. Peak flux densities with fitted sky level have the best performance. If the expected flux density is larger than 100\uJy, the measured peak flux density is accurate, with an average error equal to the noise level of the image. When the expected flux density is 50\uJy\ ($\sim 7 \sigma$), the measured peak flux density is overestimated by $\sim 7$ per cent.

We choose peak flux densities with fitted sky as our flux density measurement, appropriate since all sources (except one) are unresolved. The measured flux densities of 101 objects are present in Table~\ref{table:flux}. In cases with stronger sidelobes from nearby bright sources, we carefully monitor the fitted sky level and ensure that the fitted integrated flux density is reasonably close to the value of fitted peak flux density.  

Flux density measurements of snapshot radio data suffer from missing flux density during the cleaning process \citep{Becker1995,White1997}, so called CLEAN bias, which is not due to the undersampling resulting from incorrectly large pixel size nor the lack of spatial frequency coverage. The CLEAN bias is not well-understood, with listed values varying among different studies. \cite{White1997} show that CLEAN bias is independent of flux density of a source but dependent on the distance from the field center. In contrast, \cite{Hodge2011} find a flux density dependence in their CLEAN bias but do not see a distance dependence. In our test shown in Table~\ref{table:cleanbias}, we do not see significant missing flux density. \cite{Rau2016} attribute the CLEAN bias to crowded fields and test the performance of several cleaning modes, showing that either the use of the cleaning masks or of the MFS mode is able to eliminate the effect. In this paper, we use both cleaning masks and MFS mode, which may be the reason that we do not see a CLEAN bias and therefore we do not apply any bias correction to our flux densities. 

\subsection{Measurements of spectral indices}
\label{sec:spectral-index-method}

Depending on the flux densities, we use some of the following five methods to measure the spectral indices $\alpha$, defined as $F_\nu \propto \nu^\alpha$ and listed in Table~ \ref{table:spectral-indices}. (1) During data reduction, the MFS mode can be used with \texttt{nterms $\geq 2$} to model the radio spectrum in each pixel as a Taylor expansion in frequency, generating the $\alpha_{\rm MFS}$ map as one of the cleaning products. (2) As the observing frequency spans from 4.25 to 8.2\,GHz, the in-band spectral indices $\alpha_{\rm inband}$ can be measured by cleaning two frequency bands with a bandwidth of 2\,GHz respectively. These two methods are somewhat similar in that they take advantage of the spectral range of our VLA observations, but spectral indices between 4.25 and 8.2\,GHz for individual sources can only be obtained by methods (1) and (2) when their flux densities $F_\nu \gtrsim 150$\uJy. The advantage of method (1) is that if the object is bright enough, its entire spectral index map is available. However, in practice sometimes only the brightest pixel has an index measurement, which is therefore not very robust. The advantage of the more conventional method (2) is that we can calculate the uncertainty of spectral indices by propagating the uncertainty of flux densities. 

(3) If a source is detected in both our VLA observation and in FIRST, its spectral index between 1.4 and 6.2\,GHz can be measured. This is the case for SDSSJ1550+0806 since it is above the FIRST catalog threshold. An additional five with $F_\nu \gtrsim 150$\uJy\ at 6.2\,GHz are marginally detected in FIRST, and their spectral indices are measured using the peak flux densities in the FIRST images. There are some uncertainties in this method.
First, these sources are detected at only $\sim 4\sigma$ levels in FIRST. Furthermore, CLEAN bias is known to affect FIRST flux densities, and while we correct the flux densities from FIRST by a CLEAN bias of the lesser of 250\uJy\ or 40 per cent of the measured peak flux densities \citep{White2007}, it may have a poorly known dependence on the location of a source in an image as well as on the flux density of the source. SDSSJ2215$-$0056 is detected in the deeper 1.4\,GHz radio observations of the Stripe 82 catalog \citep{Hodge2011}, and therefore its spectral index between 1.4\,GHz and 6.2\,GHz is available. When we calculate the spectral index for SDSSJ2215$-$0056, we correct it for the CLEAN bias of 35\uJy, which is not included in the Stripe 82 source catalog. 

For fainter sources, we use the stacking methods to obtain average spectral indices. (4) By stacking VLA images in two frequency bands, we can measure the average spectral indices between 4.25 and 8.2\,GHz. To avoid heavily weighting the brighter sources, we calculate the average spectral indices $\langle \alpha \rangle$ by normalizing the contribution of each object by its flux density at the central frequency: 
\begin{equation}
\langle \alpha \rangle_{\rm inband} = \frac{\log\left(\sum\limits_i \frac{F_{i,1}}{F_{i,c}}  / \sum\limits_i \frac{F_{i,0}}{F_{i,c}} \right)}{\log(\nu_1 / \nu_0)},
\end{equation}
where $\nu_0=5.1215$\,GHz and $\nu_1 = 7.2635$\,GHz. $F_{i,c}$ is the flux density of object $i$ at the central frequency 6.2\,GHz, while $F_{0,c}$ and $F_{1,c}$ are the flux densities at $\nu_0$ and $\nu_1$ respectively. 

(5) By stacking the FIRST images, we can also measure the average spectral indices between 1.4 and 6.2\,GHz, normalizing the contribution of each object by its flux density at 6.2\,GHz:
\begin{equation}
\begin{multlined}
\langle \alpha \rangle_{\rm outband}  \\
= \frac{\log \left(\sum\limits_i \frac{F_{i,6.2}}{F_{i,6.2}} /  \sum\limits_i \frac{F_{i,1.4}}{F_{i,6.2}}\right)}{\log(\nu_{6.2} / \nu_{1.4})} 
= \frac{ \log\left( N/ \sum\limits_i \frac{F_{i,1.4}}{F_{i,6.2}}\right)}{\log(\nu_{6.2} / \nu_{1.4})},
\end{multlined}
\end{equation}
where $\nu_{1.4} = 1.4$\,GHz and $\nu_{6.2}=6.2$\,GHz. $F_{i,1.4}$ and $F_{i,6.2}$ are the flux densities of object $i$ at 1.4 and 6.2\,GHz respectively, and $N$ is the number of sources used in the stacking. In contrast to the flux density measurement of individual objects, for stacking we use the peak flux density of the brightest pixel, and the uncertainty of the average spectral indices is estimated by bootstrapping analysis. We apply a CLEAN bias of 40 per cent of the peak flux density to the FIRST measurements. For methods (4) and (5), we also measure the spectral indices without flux density normalization, and the difference in the spectral indices resulting from methods with or without weights is smaller than their reported uncertainties. 

\section{Radio properties of ERQs}
\label{sec:prop}

\subsection{Radio luminosity distribution}

While the exact shape of the radio luminosity distribution continues to be under debate, objects at the bright end of the radio luminosity function are conventionally called `radio-loud', and this part of the AGN population has long been associated with powerful jets \citep{Urry1995}. The exact dividing line between the radio-loud and the radio-quiet population is difficult to determine, and in fact may depend on the bolometric luminosity of the AGN and other properties. Traditionally, radio-to-optical ratios have been used to delineate radio-loud from radio-quiet sources, but such measures are not particularly meaningful for ERQs whose optical emission is affected by obscuration as is clear from their spectral energy distributions \citep{Zakamska2016, Hamann2017}. Another possibility is to use forbidden emission lines which are emitted on larger scales and are less affected by obscuration \citep{Xu1999}. For the typical \oiii\ luminosity of ERQs of $10^{10}$\Lsun\ \citep{Zakamska2016}, \citet{Xu1999} would consider AGNs with $\nu L_{\nu}$[5\,GHz] $\ga 10^{41.8}$\ergs\ radio-loud. 

The flux densities used for calculating the luminosity function of ERQs are listed in Table~\ref{table:flux}. Given a median noise level of $7.6$\uJy, the typical detection threshold for our observations is $4\sigma \sim 30$\uJy. At this sensitivity, radio emission is detected at 6.2\,GHz in 65 out of 101 sources in our VLA observations. Radio luminosities are computed using equation
\begin{equation}
\nu L_{\nu} = 4 \pi D^2_L (1+z)^{-1-\alpha} (\nu / \nu_{\rm obs})^{1+\alpha} \nu_{\rm obs} F_{\nu_{\rm obs}},
\end{equation}
where $D_L$ is the luminosity distance at redshift $z$. For comparison with previous studies, we $k$-correct to rest-frame $\nu=6$\,GHz, with $\nu_{\rm obs}=6.2$\,GHz for our VLA observations and $\nu_{\rm obs}=1.4$\,GHz for objects with FIRST catalog detections. We adopt a spectral index of $\alpha = -0.7$ for the $k$-correction. Although the spectral indices of brighter individual sources can be measured, we use $\alpha=-0.7$ for them in order to be consistent with fainter sources in the measurements. If instead of $-0.7$, $\alpha$ is $-1.0$, then our calculations underestimate the intrinsic luminosity at 6\,GHz by a factor of $\sim 1.5$, or $0.18$\,dex.

\begin{figure}
\centering
\includegraphics[scale=0.43]{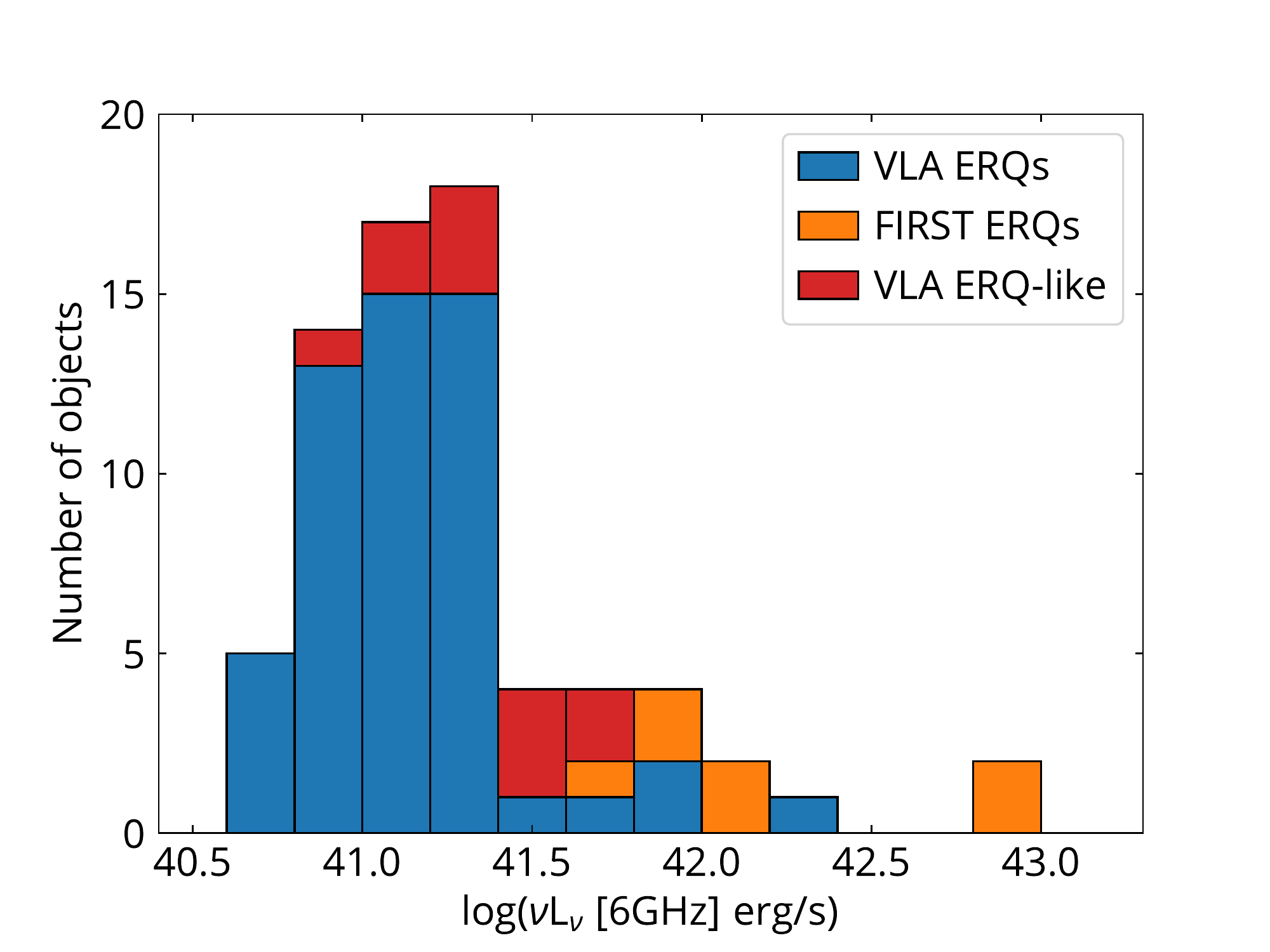} 
\includegraphics[scale=0.43]{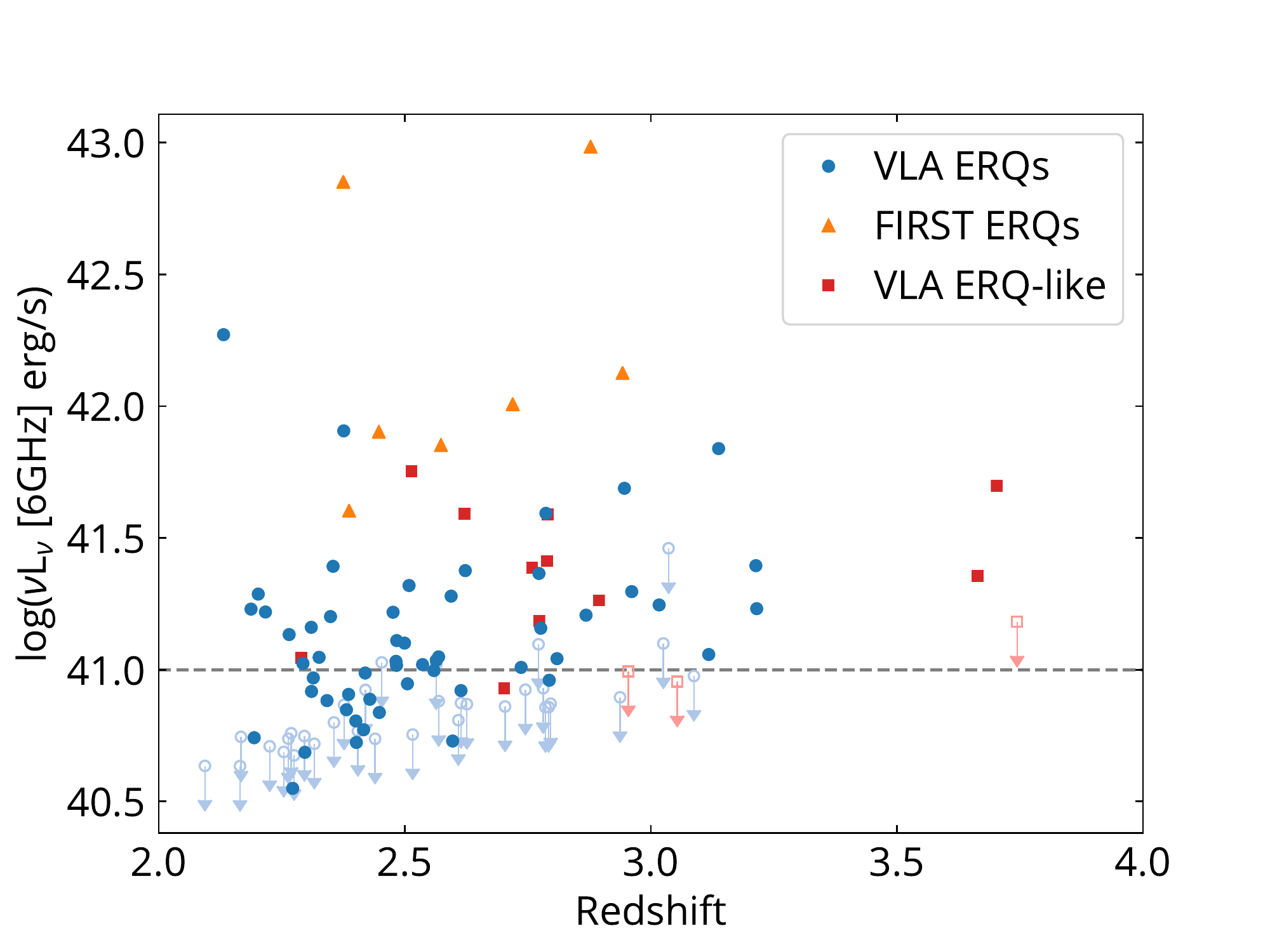}
\caption{Top: radio luminosity distribution of radio-detected ERQs $k$-corrected to rest-frame 6\,GHz. Bottom: radio luminosities versus redshifts. Blue dots are ERQs observed with VLA, orange triangles are ERQs detected in FIRST, and red squares are ERQ-like objects. The upper limits for non-detections are set to $4\sigma$, where $\sigma$ is the root-mean-square noise in the their fields. The dashed line at $10^{41}$\ergs\ indicates the luminosity above which most (90 per cent) objects are detected over the entire redshift range.}
\label{fig:LD}
\end{figure}

The distribution in radio luminosity and redshift for all ERQs in our study is shown in Figure~\ref{fig:LD}. Within the limited redshift range, there is no obvious dependence of radio activity on the redshift, except that the upper limits are tracking the typical detection limit of our observations. Our dataset is nearly complete (90 per cent, including FIRST sources) down to $\nu L_{\nu}$[6\,GHz]=$10^{41}$\ergs. The radio luminosity of detected sources spans nearly three orders of magnitude, from $10^{40.5}$ to $10^{43.0}$\ergs. The median luminosity of ERQs at 6\,GHz is $1.1\times 10^{41}$\ergs.

\begin{figure}
\centering
\includegraphics[scale=0.43]{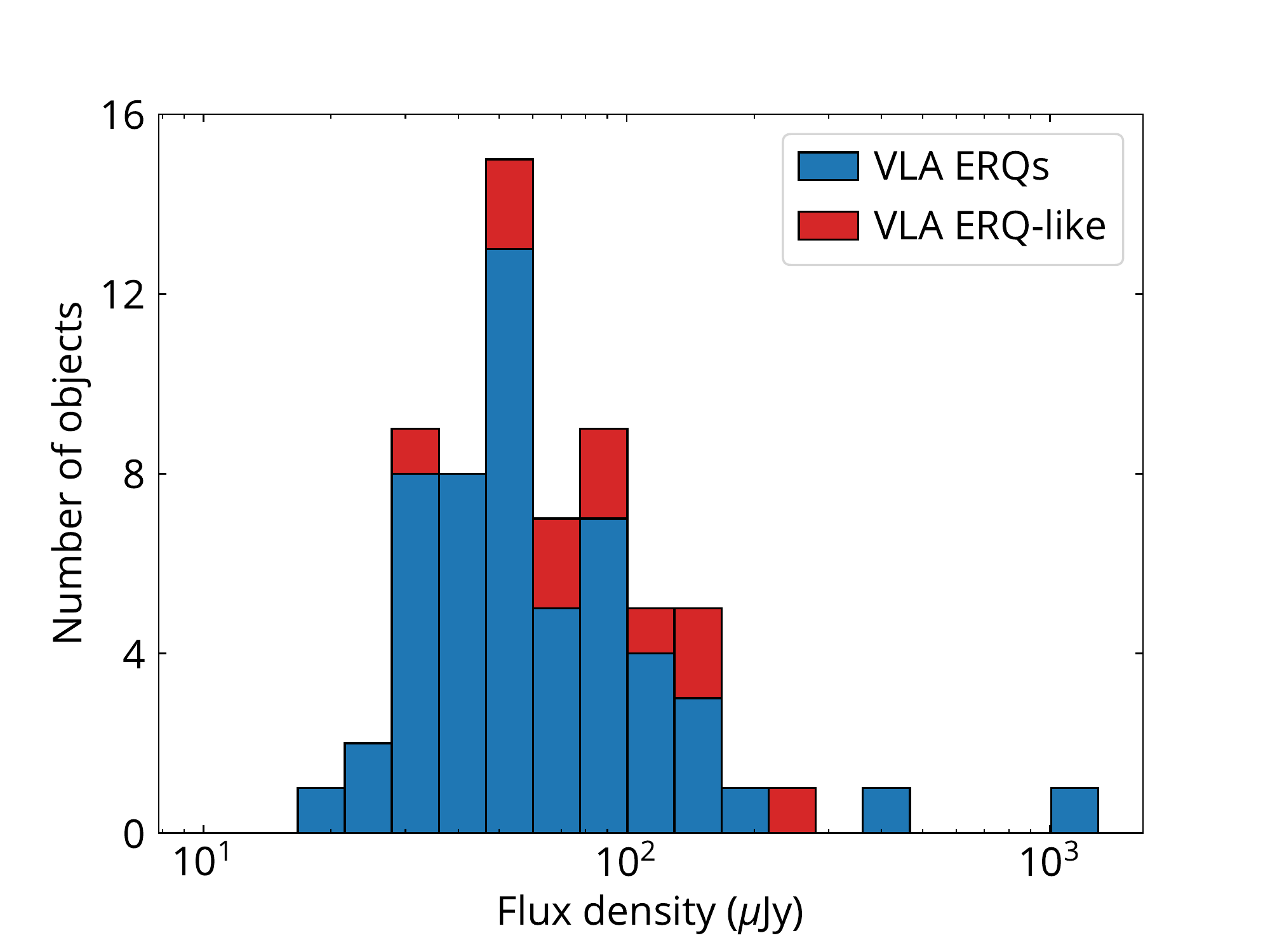} 
\caption{The distribution of flux densities observed at 6.2\,GHz for ERQs detected in the VLA observations.}
\label{fig:flux-distribution}
\end{figure}

\begin{figure*}[t]
\centering
\includegraphics[scale=0.8]{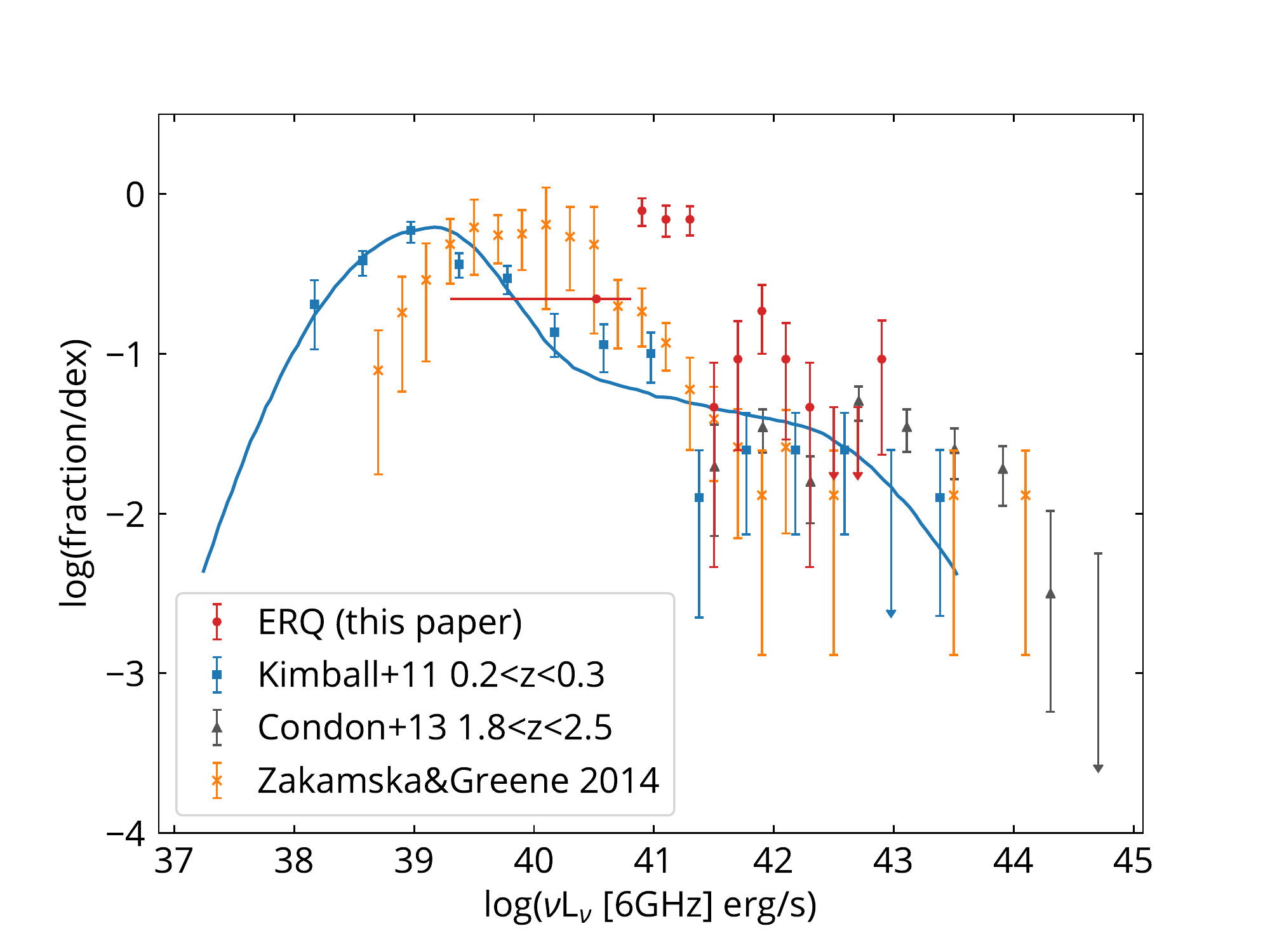} 
\caption{
The distribution of radio luminosity at rest-frame 6\,GHz, presented in the fraction of overall parent population per logarithmic bin. The red dots represent ERQs in this paper, and the red horizontal segment is the estimated radio luminosity for 36 non-detections, where the horizontal range indicates the standard deviation of the overall radio luminosity distribution for non-detections, estimated by bootstrap analysis. For comparison, the blue squares are for the low-redshift AGN sample in \cite{Kimball2011}, the orange crosses are for $z<0.8$ type 2 quasars in \cite{Zakamska2014} and the grey triangles are for high-redshift type 1 quasars in \cite{Condon2013}. The error bars of the sample from \cite{Zakamska2014} are estimated by considering the flux densities of radio non-detections to be 1, 2, and 10 times fainter than their upper limits, resulting in larger uncertainty. The radio-quiet bulk of the high-redshift type 1 quasar population is not detected by \cite{Condon2013} and is therefore not shown. Compared to other AGN samples, the radio luminosities of ERQs are concentrated around a much higher median luminosity $1.1\times10^{41}$\ergs. 
}
\label{fig:LF}
\end{figure*}

In Figure~\ref{fig:LF} we present the radio luminosity distribution of ERQs in comparison with the radio luminosity distributions of several other quasar populations. Although in general ERQs and ERQ-like objects share similar properties, we exclude ERQ-like objects in this calculation to avoid potential contamination. As the number of ERQs which are not selected for the follow-up SDSS spectroscopy and therefore are not in our sample is not well-understood, we cannot compute the volume density at each radio luminosity accurately. Instead, for every bin in the logarithm of the radio luminosity, we show the fraction of objects with radio luminosities within this bin in the targeted ERQ population. In this representation, if the luminosity function is measured over the full luminosity range, then the integral under the curve should amount to 1, and differently binned luminosity functions can be directly compared to one another. In order to avoid confusion with the luminosity function defined in volume density, we use `luminosity distribution' for the representation in Figure~\ref{fig:LF} in this paper. We emphasize that after we discard those objects targeted for follow-up spectroscopy only as radio sources, the selection for spectroscopy is independent of the radio properties.

The detection limit of the VLA observations is nearly complete (90 per cent) for radio luminosity above $10^{41}$\ergs. We perform a completeness correction for the radio luminosity bin of $10^{40.8-41.0}$\ergs\ by assuming a similar radio luminosity distribution between the redshift bin $z=2.0-2.4$ and the radio non-detections with upper limits $>10^{40.8}$\ergs. Specifically, out of 32 sources in the redshift bin $z=2.0-2.4$, 11 are detected above $10^{41}$\ergs, 3 detected in $10^{40.9-41}$\ergs, and 3 in $10^{40.8-40.9}$\ergs. We use these detection counts to determine the probabilities of detection in the radio luminosity bin $10^{40.8-41.0}$\ergs\ for the non-detections in the entire redshift range: 5 non-detections with upper limits higher than $10^{41}$\ergs, 4 with upper limits in $10^{40.9-41.0}$\ergs, and 10 with upper limits in $10^{40.8-40.9}$\ergs. This gives a completeness correction of additional four counts in the bin of $10^{40.8-41.0}$\ergs. There are already 13 detections in this luminosity bin and thus this completeness correction and its uncertainty do not change the overall shape of the ERQ radio luminosity distribution. Error bars for the ERQ luminosity distribution are estimated using bootstrap analysis.

Out of our 101 objects, 36 sources are not detected in the VLA observations. To estimate the mean flux density for non-detected objects, we exclude 5 fields where noise levels are $>10$\uJy\pbeam\ and stack the remaining 31 fields containing non-detections. The stack reveals a detection with the mean and median flux densities of $13.8\pm 1.8$\uJy\ and $14.9\pm 2.5$\uJy\ respectively, where the uncertainty is estimated by bootstrap analysis. At their median redshift $z=2.57$, the median flux density corresponds to a radio luminosity at 6\,GHz of $3.3 \times 10^{40}$\ergs. This enables us to place the lowest radio luminosity point (red horizontal segment) in Figure~\ref{fig:LF}.

The method to estimate the radio luminosity distribution for the radio non-detections from \cite{Zakamska2014} in Figure~\ref{fig:LF} is different from that used for ERQs. In the sample of \cite{Zakamska2014}, all 7 FIRST non-detections located in Stripe 82 are detected in the deeper survey \citep{Hodge2011} with flux densities between 0.4 and 0.7\,mJy, about 2 times fainter than the detection limits of FIRST. Based on this, the radio luminosity distribution for the sample of \cite{Zakamska2014} is derived by considering the flux densities of non-detections to be 1, 2, and 10 times fainter than their upper limits. In each case, error bars are estimated using bootstrap analysis, and the error bars in Figure~\ref{fig:LF} show the largest ones among these three cases. This calculation results in larger uncertainties when the radio luminosity is smaller than $10^{40.5}$\ergs, while those bins above this luminosity are well constrained by the detections. The FIRST stacking of these non-detections has a median of $301\pm 24$\uJy, without a correction for the CLEAN bias. After the CLEAN bias correction, the median radio luminosity is $4.6\times10^{39}$\ergs, consistent with the radio distribution of \cite{Zakamska2014} in Figure~\ref{fig:LF}.

\subsection{Morphologies and spectral indices}
\label{sec:indices}

Typically, in radio-loud AGNs flat spectral indices ($\alpha>-0.5$) are due to relativistic boosting of self-absorbed synchrotron spectra arising in a compact $\la 1$ pc core, whereas steep spectral indices ($\alpha<-0.5$) are due to unboosted radio emission arising on larger scales \citep{Blandford1979, Orr1982}. Spectral indices of radio-quiet AGNs are still not well known due to the lack of a sensitive all-sky radio survey at a different frequency from FIRST.

Spectral indices measured by methods described in Section~\ref{sec:spectral-index-method} are listed in Table~\ref{table:spectral-indices}. Our inband spectral indices correspond to rest-frame frequencies $\sim 15-30$\,GHz, and our outband spectral indices correspond to rest-frame $\sim 5-20$\,GHz. Because the observed radio spectra are steep, free-free emission, which has a spectral index of $-0.1$, cannot dominate the radio emission of ERQs at these frequencies. In order to get measured $\alpha_{\rm inband}=-1.0$, the fraction of free-free emission is limited to $\sim$40 per cent at rest-frame $\sim20$\,GHz, assuming that the underlying synchrotron spectrum cannot be steeper than $\alpha=-1.5$. This fraction should be considered as a conservative upper limit because we do not see significantly flatter $\alpha_{\rm inband}$ compared to $\alpha_{\rm outband}$, which is expected if free-free emission dominates at higher frequencies. At face value, this upper limit on the free-free emission suggests that the star formation rates in ERQ hosts must be limited to $\sim 2000$\Msunyr \citep{Murphy2012}. This conversion is uncertain because the existing scalings between free-free emission and star formation have not been calibrated in this radio luminosity regime.

ERQ SDSSJ1117+4623 is the brightest object (1.25\,mJy at 6.2\,GHz, or $\nu L_{\nu}$[6GHz]$=1.9\times 10^{42}$\ergs) in our VLA observations. It is also marginally detected in FIRST with the brightest pixel having an intensity of 0.73\,mJy\pbeam, though it is not bright enough to be listed in the FIRST catalog. Its spectral index between rest-frame 12\, and 25\,GHz is $+0.29\pm0.15$ (MFS mode) or alternatively $+0.37\pm0.09$ (two-band imaging), while the spectral index between rest-frame 4 and 19\,GHz is $+0.17\pm0.10$, corrected for the CLEAN bias of FIRST. These measurements are consistent with one another within the uncertainties. Among the sources with measurable spectral indices, SDSSJ1117+4623 is the only one with a flat spectral index of $\alpha>-0.5$. Interestingly, while the main source itself is not spatially resolved, there is a marginally resolved faint radio source with a peak intensity of 55\uJy\pbeam\ and an integrated flux density of 90\uJy\ to the south-east of SDSSJ1117+4623 at a projected distance of 80\,kpc (Figure \ref{fig:image-example}). This object lacks an SDSS counterpart, and we hypothesize that SDSSJ1117+4623 consists of a flat-spectrum core and a lobe with a hotspot. Therefore, the radio emission from this source may be due to a jet. However, it is still possible that the extended radio source is from an unrelated source. For example, dust strongly attenuates UV and optical light but does not affect radio wavelengths, resulting in the non-detection in SDSS.

Aside from SDSSJ1117+4623, all other sources with individually measured spectral indices have steep radio spectra with $\alpha$ ranging from $-1.8$ to $-0.7$ and they are all point-like -- the fitted 2D elliptical shapes we use for measuring flux densities are consistent with the theoretical PSF provided by CASA. The stacking analysis of fainter sources without individual indices also yields steep spectra. The brighter flux density bin (80 $-$ 150\uJy) shows an average spectrum with $\alpha=-1.08\pm 0.25$, and the fainter flux density bin (30 $-$ 80\uJy) shows an even steeper average spectrum with $\alpha=-1.35 \pm 0.34$.

The stacks reveal extended emission between 2 and 3\,arcsec from the center of the synthesized beam, inconsistent with the expectations from the Gaussian beam shape and with an average intensity excess of $\sim 2$\uJy\pbeam. Because the flux density of the extended emission is sensitive to the cleaning threshold, we suspect this is a cleaning artifact, with residual PSF below the cleaning threshold. 

\section{Discussion}
\label{sec:disc}

\subsection{Summary of observed ERQ properties}

In the previous section, we find that, as a population, ERQs have the following radio properties: 
\begin{itemize}
\item At rest-frame 6\,GHz, 9 ERQs ($\sim 8$ per cent of the population) have radio luminosities $\nu L_{\nu}$ in excess of $10^{41.8}$\ergs, the nominal `radio-loud' definition we adopt for this population based on the \oiii\ luminosity ($\log L_{\rm 5GHz} = 24$\,W\,Hz$^{-1}$Sr$^{-1}$ in \citealt{Xu1999}). This fraction is qualitatively similar to the radio-loud fractions of other quasar subpopulations, though it depends sensitively on the exact definition. Since ERQs are candidate high-redshift obscured quasars, we compare their radio-loud fraction to that of low-redshift type 2 quasars and find that the radio-loud fraction is similar ($\sim 8$ per cent; \citealt{Zakamska2004a}) when the \oiii-based radio-loudness definition is adopted.
\item The median 6\,GHz radio luminosity is $1.1\times 10^{41}$\ergs, with $\sim 78$ per cent of the population with luminosities between $10^{40.0}$ and $10^{42.0}$\ergs. This median radio luminosity is $\sim1.5$\,dex higher than the low-redshift sample from \cite{Zakamska2014}, and $\sim0.5$\,dex higher than the sample from \cite{Condon2013}. 
\item We find no sources extended on $\ga$10\,kpc scales except for one powerful source SDSSJ1117+4623 on the radio-loud end of the distribution, which can be explained as a jet core+lobe system.
\item With the exception of the same core+lobe system which has a flat spectrum core, ERQs have steep radio spectra with $\langle \alpha \rangle$ between $-1.0$ and $-1.3$ depending on the measurement method and a standard deviation within the population of $0.3$.  
\end{itemize}

As ERQs were selected by their high mid-infrared-to-optical ratios, it is not surprising that their spectral energy distributions steeply rise from optical to infrared wavelengths \citep{Zakamska2016, Hamann2017}. Dust obscuration with $A_{\rm V}$ of a few to 10\,mag can account for the shape of the near-infrared spectral energy distribution, but not for the rest-frame ultra-violet continuum which is too strong and too blue to be directly visible reddened light from the quasar \citep{Hamann2017}. \citet{Alexandroff2017} demonstrate that the rest-frame ultraviolet continuum is highly polarized and is therefore largely due to scattered light, as expected in the classical geometric unification model of AGNs \citep{Antonucci1993} and as seen in low-redshift powerful type 2 quasars \citep{Zakamska2006, Greene2012, Obied2016}. Goulding et al. in prep. find that ERQs are high-luminosity ($L_{\rm X}\ga 10^{45}$\ergs), strongly obscured ($N_{\rm H}\simeq 10^{24}$\,cm$^{-2}$) X-ray emitters. In summary, ERQs show multi-wavelength properties consistent with a luminous, obscured quasar population, where the powerful infrared emission is produced by warm dust near the quasar. In particular, the spectral energy distribution is inconsistent with BL Lac-type synchrotron-dominated spectra \citep{Fossati1998} which have flatter radio indices with no strong variations in the spectral energy distribution as a function of wavelength analogous to the steep rise seen in ERQs toward infrared wavelengths, so we do not consider this possibility further. 

Follow-up near-infrared spectroscopy reveals outflows of ionized gas as traced by the \oiii\ emission with unprecedented kinematics (\citealt{Zakamska2016}, Perrotta et al. in prep). The ERQs' infrared luminosities (obtained from the apparent $W3$ and $W4$ fluxes) at rest-frame $5$\,$\mu$m range from $10^{46-47}$\ergs\ \citep{Zakamska2016, Hamann2017}, with a median of $10^{46.4}$\ergs\ for the objects in our radio sample. To estimate the bolometric luminosity, we use a bolometric correction factor of 8 \citep{Richards2006} to arrive at the median bolometric luminosity of $L_{\rm bol}=10^{47.3}$\ergs. As the Eddington mass for a luminosity of $10^{47}$\ergs\ is $\sim 10^9$\Msun, ERQs might be accreting at near- or even super-Eddington levels, which may explain their unusually strong outflow activity. Given this context for multi-wavelength properties of ERQs, in the following sections we discuss the origin of the radio emission from ERQs. 

\subsection{Is radio emission from ERQs due to star formation?}
\label{sec:not-SF}
Radio synchrotron emission is a known by-product of star formation, as supernova explosions shock the interstellar medium, enriching the galaxy with relativistic particles. This process results in a strong correlation between far-infrared luminosities and radio luminosities in star-forming galaxies \citep{Helou1985}. The median ERQ radio luminosity of $\nu L_{\nu}$[6\,GHz]$=1.1\times 10^{41}$\ergs\ would require a star formation rate of 2700\Msunyr\ \citep{Bell2003}. This is an extremely high value, but it does not contradict the observed infrared fluxes from \wise. By scaling the templates of spectral energy distributions of star-forming galaxies \citep{Rieke2009} to these star formation rates, we find that the $W4$ fluxes expected from star formation of this level are 4 magnitudes fainter than observed, indicating that the \wise\ fluxes of ERQs are dominated by quasar activity, not star formation.

Such values of star formation are not found in the local Universe, and they are at the upper limit of values found in the most luminous submillimeter galaxies at ERQ redshifts \citep{Blain2002, Chapman2005}. Is it possible that ERQ host galaxies support these levels of star formation required to explain the observed radio emission? While there are not yet far-infrared measurements of star formation available for ERQs, some authors have argued that the star formation in a closely related population -- hot dust obscured galaxies (HotDOGs; \citealt{Fan2016b, Farrah2017}) -- can contribute 10$-$20 per cent of the bolometric luminosity, which at $L_{\rm bol}=10^{47.3}$\ergs\ would translate into 900$-$1800 \Msunyr \citep{Bell2003}. These are also extremely high values, but they are smaller than is required to power the median radio luminosity in our sample. Therefore, the top 50 per cent of the radio luminosities of the radio-quiet ERQs ($10^{41}-10^{42}$\ergs) are hard to reconcile with star formation.

Another puzzle is the steep radio spectra of ERQs. In local starburst galaxies, the radio spectral indices at $\sim$GHz frequencies are expected to be $-0.7$, flattening both at higher frequencies (10$-$100\,GHz) where free-free emission contributes \citep{Murphy2011, Murphy2012} and at lower frequencies where in compact starbursts self-absorption effects become important \citep{Murphy2013}. In contrast, the mean spectral index of ERQs is $-1.0$ with one source showing $\alpha=-1.8$, significantly different from the expected values of starburst galaxies. \cite{Ibar2010} study the radio-identified submillimeter galaxies at $z\sim 2$ and measure a mean spectral index between 610 MHz and 1.4\,GHz of $-0.75\pm0.06$, indistinguishable from those of local starburst galaxies and appreciably higher than those of ERQs. A comparison of ERQs with this population suggests that the radio emission in ERQs is not powered by star formation.

In starburst galaxies, inverse Compton scattering might accelerate cooling and steepen the synchrotron spectra \citep{Seaquist1991}. However, this process would also suppress radio emission, making it even more difficult to reproduce the observed radio emission with star formation in ERQs. Thus on the balance of available data we find it unlikely that star formation dominates the radio luminosity of ERQs. The required star formation rates are too high on the upper end of the radio-quiet subset of ERQs ($\nu L_{\nu}\la 10^{42}$\ergs), and the spectral indices are steeper than those that are found both in local starbursts and in high-redshift submillimeter galaxies.

\subsection{Shape of the radio luminosity distribution}

In Figure~\ref{fig:LF}, we compare the radio luminosity distribution of ERQs with three other datasets: nearby type 1 quasars \citep{Kimball2011}, intermediate redshift type 2 quasars \citep{Reyes2008, Zakamska2014} and high-redshift type 1 quasars \citep{Condon2013}. All of the observations are based on optically selected sample, and they have been corrected to rest-frame 6\,GHz here. The most complete coverage of the radio luminosity function was achieved by \citet{Kimball2011} who detect 173 out of 179 quasars in their sample at $0.2<z<0.3$ and fit a two-population model to their observed radio luminosity function. By comparing their data to the radio luminosity functions of star-forming galaxies and AGNs, they suggest that the peak centered at $10^{39}$\ergs\ is due to star formation, whereas AGNs dominate radio output at brighter luminosities. The star formation rate of $25$\,\Msunyr\ required to explain the median radio luminosity is high but not unreasonable. 

A similar argument was put forward by \citet{Condon2013} who present the radio luminosity function of 2471 type 1 quasars at $1.8 < z < 2.5$. Only 8 per cent -- the radio-loud end of the population -- are detected at $F_\nu > 2.4$\,mJy by the National Radio Astronomy Observatory VLA Sky Survey \citep[NVSS;][]{Condon1998}, and those are the objects shown with data points in Figure \ref{fig:LF}. By analyzing the statistics of the radio maps of undetected sources, these authors suggest that there is a peak of the luminosity function at faint luminosities with an estimated median radio luminosity $\nu L_{ \nu}=10^{40.4}$\ergs, which we have translated to rest-frame 6\,GHz for direct comparison with our data assuming $\alpha = -0.7$. By stacking radio data, \cite{White2007} find similar mean radio luminosities for the most optically luminous quasars in their sample. \cite{Condon2013} argue that this peak may be due to powerful star formation in the quasar hosts at the level of $\sim 500$ \Msunyr, though no independent star formation measures for this population are available to probe this hypothesis. 

We also present the radio luminosity function of $z<0.8$ type 2 quasars from the sample of \citet{Zakamska2014}, where to avoid biasing our measurements to high radio luminosities we have excluded all objects targeted for follow-up SDSS spectroscopy only due to their radio emission. The median radio luminosity of this sample, translated again to rest-frame 6\,GHz, is $10^{40.2}$\ergs, which would correspond to a star formation rate of $\sim 300$\Msunyr\ \citep{Bell2003}. However, for this sample several multi-wavelength measures of star formation are available and are strongly inconsistent with the star formation rates implied by the observed radio luminosity, so that only $\sim 10$ per cent of the observed radio luminosity can be due to star formation \citep{Zakamska2016b}. In other words, another mechanism is required to explain the radio emission of powerful type 2 quasars. As these objects are unresolved at 5 arcsec ($\sim 30$\,kpc) scales in the FIRST survey, there is no evidence that their radio emission is due to large-scale jets. 

To summarize, even for a narrow range of bolometric luminosities or optical properties of AGNs, the radio luminosity distribution can span over six orders of magnitude. The radio luminosity cannot fall below that associated with star formation in the host galaxy, and this star formation component can dominate the lower end of the radio luminosity function \citep{Kimball2011, Rosario2013}. On the `radio-loud' end, we have a population that has long been associated with powerful jets \citep{Urry1995}. The nature of the radio emission in the intermediate radio luminosity $10^{39}-10^{42}$\ergs\ range is surprisingly ambiguous, and this is exactly the regime in which the radio luminosity distribution of ERQs peaks. Furthermore, we find that the peak of the luminosity distribution of ERQs occurs at appreciably higher radio luminosities than the peaks for the comparison samples, while ERQs still remain predominantly radio-quiet. 

Various possible mechanisms have been discussed to explain the radio output of AGNs in this intermediate luminosity range. One possibility is coronal emission around the accretion disks \citep{Laor2008, Raginski2016}. Because this emission arises on scales $\la 10^5$ Schwarzschild radii, it is expected to be optically thick to synchrotron self-absorption at the frequencies of our observations and therefore have flat spectra ($\alpha=0$). As ERQs have steep radio spectra, we rule out this mechanism as the dominant origin of their radio emission. 

In Figure~\ref{fig:LF}, the radio luminosity distributions of ERQs and of quasars from \citet{Kimball2011} and \citet{Condon2013} show a local minimum at $\sim 10^{41.4}$\ergs. Interestingly, this minimum appears in the radio luminosity functions of quasars of widely different bolometric luminosities, which is in contradiction with a luminosity-dependent dearth of radio sources seen by \citet{Xu1999}. A local minimum at $\sim 10^{41.4}$\ergs\ in the luminosity function suggests that different mechanisms might be responsible for the radio emission below and above this value.

We now proceed to comparing the features in the luminosity distributions shown in Figure~\ref{fig:LF} to the traditional divides between radio-loud and radio-quiet objects. Because ERQs are obscured at optical wavelengths, we use the rest-frame infrared luminosity at 5\um\ as an indicator of bolometric luminosity (Goulding et al., submitted). As a comparison sample, we use type 1 quasars in a narrow redshift range $0.5<z<0.7$ from DR12Q \citep{Paris2017} so as to avoid introducing an artificial correlation in the infrared versus radio luminosity space. The infrared luminosity at rest-frame 5\um\ is obtained by power-law interpolating between \wise\ fluxes, if detected with a signal-to-noise ratio above 3. Sources targeted for follow-up SDSS spectroscopy only because of their radio emission are excluded, and the radio luminosity is corrected to rest-frame 6\,GHz assuming $\alpha=-0.7$. The FIRST-matched sources constitute 4 per cent of the \wise-detected DR12Q sample in $0.5<z<0.7$, which is a typical FIRST-detection rate in SDSS quasars and is smaller compared to that of more luminous quasars like \cite{Zakamska2014}.

\begin{figure}[t]
\centering
\includegraphics[scale=0.43]{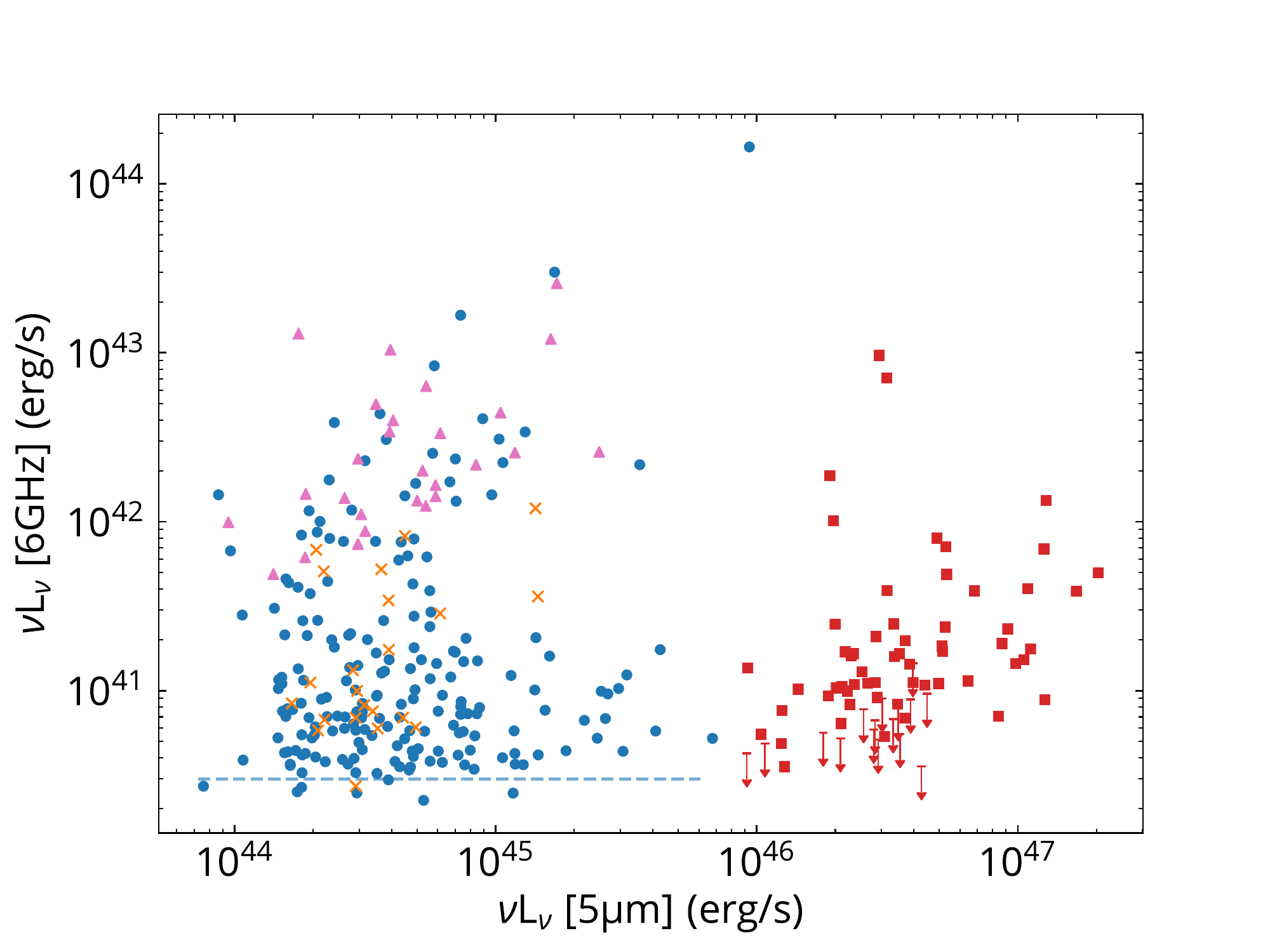} 
\caption{
Radio luminosity at 6\,GHz versus infrared luminosity at 5\,\um. The red squares represent ERQs, including ERQ-like objects and those detected in FIRST. Other symbols are quasars with redshifts $0.5<z<0.7$ from DR12Q \citep{Paris2017}. The blue points adopt the radio flux densities by cross-matching with FIRST, while the pink triangles use the ones from \cite{VanVelzen2015} to account for the emission from lobes. We visually inspect all sources, and use orange crosses to indicate jet candidates whose radio luminosities may be underestimated. The dashed line is the radio luminosity above which the low-redshift sample is complete over its redshift range.
}
\label{fig:IR_radio}
\end{figure}

The FIRST flux densities of many quasars in DR12Q are underestimated because the radio emission of lobes is not taken into account. To improve the radio measurement, we cross-match the sample with a catalog of Fanaroff-Riley type II (FR II) sources in FIRST \citep{VanVelzen2015}, and sum up the flux densities from cores and lobes. Changing the assumed spectral index from $-0.7$ to $-0.3$ would increase the calculated luminosities by 0.17\,dex. Because the automated search for FR II sources is biased to brighter sources, it might result in an artificial separation around the radio luminosity limit for the automated search. To ensure that it is not the case for the separation seen in Figure~\ref{fig:IR_radio}, we visually inspect the FIRST cutout images and mark those potential jets missed in the FR II catalog. Even if their radio luminosities are underestimated, their lower infrared luminosities show that the separation in Figure~\ref{fig:IR_radio} is not artificial.

The local minimum in the ERQ radio distribution at $10^{41.4}$\ergs\ is consistent with the separation between the radio-loud and radio-quiet population among type 1 quasars. Therefore, we conclude that ERQs are a predominantly radio-quiet population. The radio luminosity distribution of type 1 quasars with FR II morphology is markedly different from their overall luminosity distribution and is heavily biased toward higher radio luminosities. Therefore, we speculate that the traditional boundary separating radio-loud versus radio-quiet objects (which appears to happen at $10^{41.4}$\ergs\ in both populations in Figure~\ref{fig:IR_radio}) corresponds to the transition from the jet-dominated population to other mechanisms. Further measurements of radio luminosity functions, of radio morphologies and of radio spectral indices for other AGN populations around this critical luminosity will clarify the nature of this feature. 

\subsection{Radio emission and ionized gas kinematics}

In the previous sections, we argued against star formation and ruled out large-scale jets and coronal emission as the origin of the radio emission of radio-quiet ERQs. In the following sections, we consider the remaining two mechanisms -- shocks caused by wide-angle quasar winds \citep{Zakamska2014,Nims2015} and compact jets launched by central supermassive black holes \citep{Falcke2004, Leipski2006}. 

ERQs are extremely powerful quasars, which likely accrete at near-Eddington or even super-Eddington rates. On nuclear scales, such objects can produce powerful outflows via a variety of mechanisms. Radiatively driven winds in which photons are absorbed by partially ionized gas \citep{Murray1995, Proga2000, Proga2004} are a strong possibility given that ERQs show signatures of absorption-line troughs in their SDSS spectra \citep{Hamann2017}. Winds launched by radiation pressure on dust \citep{Murray2005, Thompson2015} have been recently suggested specifically to explain properties of powerful red quasars at high redshifts \citep{Ishibashi2017}. Mechanisms that involve magnetic fields are also possible \citep{Konigl1994, Everett2005}.

As these winds propagate into the surrounding interstellar medium and interact with it, the resulting shocks can produce relativistic particles which in turn yield synchrotron emission \citep{Stocke1992}. Indeed, there are indications that radio emission with luminosity $10^{39}-10^{42}$\ergs\ is related to outflow activity on galactic scales \citep{Mullaney2013, Zakamska2014}. Specifically, \citet{Zakamska2014} use low-redshift type 2 quasars to find a correlation between radio luminosity and the velocity width of the \oiii\ emission line \citep{Heckman1981}, a forbidden transition which arises in regions with density below the critical ($8\times 10^5$ cm$^{-3}$) and is now seen by multiple groups in outflow on several\,kpc scales in integral field unit observations \citep{Liu2013, Carniani2015, Rupke2017}. 

\begin{figure*}[t]
\centering
\includegraphics[scale=0.8]{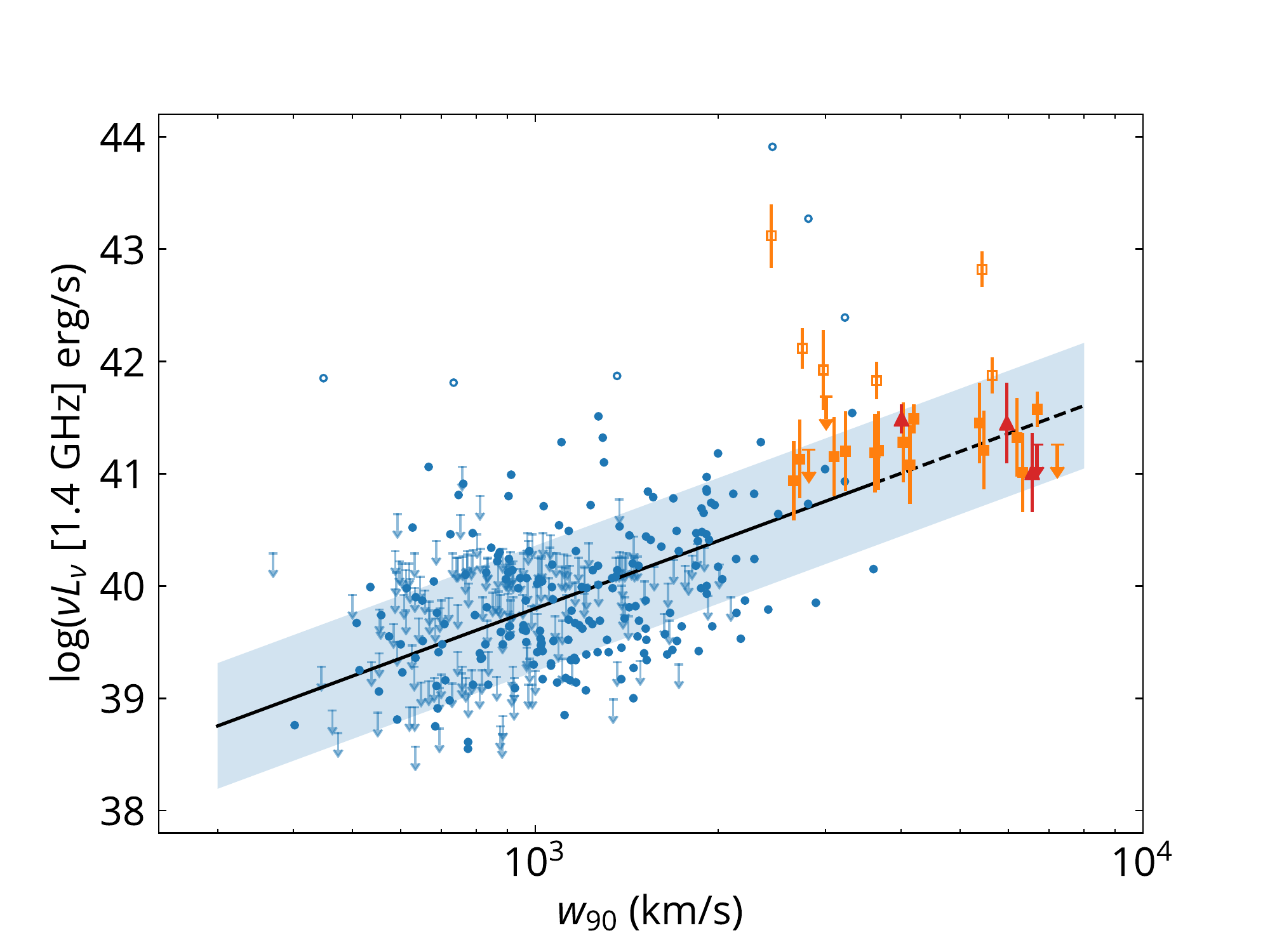} 
\caption{
Radio luminosity at 1.4\,GHz versus the velocity width $w_{90}$ of the \oiii\ emission for radio-quiet sample. The blue dots are from the low-redshift $z<0.8$ type 2 quasars \citep{Zakamska2014}, with radio luminosity upper limits of $\sim 5\sigma$ from FIRST and NVSS. The red triangles \citep{Zakamska2016} and orange squares (Perrotta et al. in prep) represent ERQs with existing \oiii\ observations. Their redshifts range from $z=2.36-2.94$ and the upper limits of radio luminosity are $4\sigma$. For ERQs without individual spectral indices, we use $\alpha=-1$ with error bars obtained by assuming $\alpha=-0.7$ and $\alpha=-1.3$. The black line is the quadratic fit derived based exclusively on the low-redshift radio-quiet sample (blue solid dots), and the shaded region represents the root-mean-square scatter. The dashed extension is to show that ERQs are not used for fitting. ERQs are located at the extreme end of both \oiii\ kinematics and radio luminosity and closely follow the correlation established for low-redshift type 2 quasars. Open symbols designate candidate radio-loud sources.
}
\label{fig:oIII_radio}
\end{figure*}

In this section, we test the hypothesis that the radio emission in ERQs is produced in a way similar to that in type 2 quasars. To this end, we check whether ERQs follow the correlation established for low-redshift type 2 quasars \citep{Zakamska2014}. For these objects, the relationship between radio luminosity and $w_{90}$, the width (in\kms) of the 90 per cent of the \oiii\ line profile power, is shown with blue dots in Figure~\ref{fig:oIII_radio} where again we have removed all objects targeted for SDSS spectroscopy only because of their radio emission. For wide-angle outflows, $w_{90}$ is a good proxy for outflow velocity and is roughly two times higher than the median radial velocity of outflowing material. \citet{Zakamska2014} suggest that the relationship is well-fit by $L_{\rm radio}\propto w_{90}^2$, implying a linear correlation between the radio power and the kinetic energy of the outflow, which may be naturally expected if a fixed fraction of the outflow energy is transferred into relativistic particles. With the fixed quadratic slope and with over half of the sample detected, the median relationship of radio-quiet sources is well determined: 
\begin{equation}
\begin{multlined}
\log(\nu L_\nu [1.4\ {\rm\,GHz}],\ {\rm erg\,s^{-1}}) \\
= 2 \times \log( w_{90}, {\rm km\,s^{-1}})  + 33.80,
\end{multlined}
\end{equation}
and the shaded region shows the root-mean-square spread of 0.56\,dex, which encompasses 65 per cent (128/196) of the detected radio-quiet type 2 quasars. Including radio-loud type 2 quasars (with radio luminosities $>10^{41.6}$\ergs\ at 1.4\,GHz based on \citealt{Xu1999}) has little effect on the fitting; it changes the intercept from 33.80 to 33.84 and results in a larger scatter (0.69\,dex).

We have been conducting near-infrared spectroscopic follow-up of the ERQs (\citealt{Zakamska2016}; Perrotta et al. in prep), and, as of this writing, 27 \oiii\ measurements are available and are shown in Figure \ref{fig:oIII_radio}. As is pointed out by \citet{Zakamska2016}, ERQs show unprecedented outflow velocities as traced by the \oiii\ emission and lie on the extreme end of the $w_{90}$ distribution. However, it is currently not known which, if any, of the ERQ properties and its selection criteria are particularly predictive of the presence of the high-velocity \oiii\ outflows. While values of $w_{90}$ in excess of 3000\kms\ are almost never seen in low-redshift samples, ERQs show an excellent agreement with the extrapolation of the correlation from \citet{Zakamska2014}: all radio-quiet radio-detected ERQs lie in the shaded region, with a root-mean-square scatter of $0.24$\,dex. While the dynamic range in $w_{90}$ probed by ERQs is too small to see the correlation within this population alone, the excellent agreement with the correlation at low redshifts strongly suggests a physical connection between radio emission and ionized gas kinematics.

\subsection{Winds as the origin of the radio emission in ERQs}

Multiple authors have discussed radio emission associated with particles accelerated on shocks which propagate through the host galaxy as a result of quasar-driven winds (e.g., \citealt{Jiang2010, Nims2015}). For quasar bolometric luminosities $10^{45}-10^{46}$\ergs, the radio luminosities predicted by these models are $\sim 10^{40}$\ergs, in the range suggested by the observations of type 2 quasars. Scaling this up to ERQ luminosities suggests that this process might naturally explain the radio luminosities of ERQs. \cite{Jiang2010} also point out that the spectral index at 5\,GHz can steepen to $\alpha<-0.8$ in $<0.3$\,Myr with the observable synchrotron lifetime of $\sim 5$\,Myr due to radiative and adiabatic cooling, which may explain the steep radio spectra in ERQs.

The existing models are idealized, with a spherically symmetric, smooth distribution of the interstellar medium. To further explore the wind hypothesis, we adopt a phenomenological approach to estimating a plausible radio luminosity resulting from quasar-driven winds, following \citet{Zakamska2014}. We assume that ERQs are capable of converting a fraction $\eta$ of their bolometric luminosity into the kinetic energy of the outflow. Then we assume that the fraction of energy that a strong shock can transfer to synchrotron-emitting particles is independent of the nature of that shock \citep{Spitkovsky2008} and take a lesson from starburst galaxies, in which $5.6\times 10^{-5}$ of the available mechanical energy supplied by stellar outflows and supernovae is transferred to the 6\,GHz radio luminosity \citep{Leitherer1999, Bell2003, Zakamska2014}. This conversion factor is the heart of the strong correlation between the far-infrared emission (which measures star formation rate) and the radio emission in star-forming galaxies \citep{Helou1985}, but now we are considering the kinetic energy of the outflow powered by the quasar instead.

The median bolometric luminosity of ERQs is $10^{47.3}$\ergs, so that the median kinetic power of their outflows is $\eta 10^{47.3}$\ergs. Applying the conversion factor from kinetic power to radio luminosity, we expect a median radio luminosity of $1.1\eta \times 10^{43}$\ergs. The observed median radio luminosity of ERQs is $1.1\times 10^{41}$\ergs\ would then require that $\eta=1$ per cent of the bolometric luminosity be converted to the kinetic power of winds. This value is in the range of the bolometric-to-kinetic conversion efficiency of quasar-driven outflows deduced from ionized gas \citep{Liu2013, Rupke2017, Sun2017} and molecular gas \citep{Fiore2017} phases, and is somewhat smaller than the bolometric-to-thermal conversion inferred from the Sunyaev-Zeldovich effect ($>10$ per cent from \citealt{Crichton2016}). If the observed radio emission is suppressed by rapid cooling \citep{Jiang2010} or inverse Compton losses \citep{Nims2015}, then the value of $\eta$ we infer by applying the efficiency of radio emission taken from star-forming galaxies should be regarded as a lower limit. 

To summarize, the wind scenario is consistent with the observed radio properties of ERQs. The radio luminosities of ERQs concentrate around the median luminosity of $1.1\times 10^{41}$\ergs, an intermediate regime where radio luminosities are too high to be explained by star formation but too low to be characteristic of classical powerful jets. The correlation between \oiii\ line width and radio emission implies a connection between the wind and the observed synchrotron radiation, and we suggest that the synchrotron-emitting particles are accelerated as the wind shocks the interstellar material of the host galaxy. The observed radio emission can be explained by winds if $\sim 1$ per cent of the bolometric luminosity is converted to the kinetic energy of wind, which is consistent with other available estimates. The steepness of the observed radio spectra is puzzling, but may be explained by rapid cooling, which is supported by theoretical work \citep{Jiang2010}. The unresolved radio morphology at $\sim 10$\,kpc scales is consistent with winds confined to their host galaxies; this emission may be resolvable in higher-resolution observations \citep{Harrison2015, Alexandroff2016}.

\subsection{Compact jets as the origin of ERQ radio emission}

Since all ERQs except for one are unresolved on scales $\sim10$\,kpc, large-scale radio jets do not dominate the radio emission of ERQs. We can also rule out core-dominated sub-parsec jets which, like radio coronae, would appear with flat spectra due to synchrotron self-absorption. However, there is a known class of jet-dominated objects -- compact steep-spectrum sources -- which are young enough to be confined to the host galaxy by interactions with the galactic gas, but have grown to sizes above the synchrotron self-absorption limit \citep{Fanti1990, Fanti1995}. In our observations at resolution $\sim 10$\,kpc, they would be seen as point-like steep-spectrum sources, as is indeed the case.

The main difference between compact steep spectrum sources and ERQs is that the former are in the unambiguously radio-loud regime, with radio luminosities between $10^{42}$ and $10^{44}$\ergs\ \citep{Holt2008}, two orders of magnitude larger than those of ERQs. One might argue that this difference reflects the selection of compact steep spectrum sources from radio surveys and hypothesize that ERQs are dominated by the lower power compact steep spectrum sources that had not been previously identified. We consider this possibility less likely on the basis of the radio distribution of ERQs. Many authors find that the radio properties of AGN populations are uncorrelated or weakly correlated with their optical properties \citep{Ivezic2002}, which is reflected in the six orders of magnitude spread of radio luminosities within each subpopulation shown in Figure \ref{fig:LF}. It seems less likely that the optical and infrared selection criteria of ERQs would pick up weak compact steep spectrum jets so efficiently that they dominate the ERQ radio luminosity function while being subdominant in every other population, especially because low-power jets are not expected to be the dominant mode of outflow activity in near-Eddington sources \citep{Jiang2014}. Nonetheless we cannot completely exclude this possibility. The physical properties of ERQs are quite extreme and poorly understood, and perhaps the ERQ selection might somehow be biased toward compact jets. For example, high mid-infrared-to-optical ratio might tend to pick objects with extremely high gas and dust content, resulting in efficient confinement of jets and therefore hitting the `sweet spot' for compact jet selection.

Another problem with the compact jet hypothesis is that the known compact jets do not fit in the tight correlation between radio luminosity and velocity width of ionized gas in Figure~\ref{fig:oIII_radio}. Compact steep spectrum sources are known to be associated with high-velocity \oiii\ outflows produced as the jet interacts with the interstellar medium of the host galaxy. While the compact radio sources in \cite{Holt2008} are two orders of magnitude brighter in radio than ERQs, the full width at half maximum (FWHM) of \oiii\ spans widely from $\sim300$ to $1500$\kms. In contrast, ERQs with existing \oiii\ observations frequently have high velocity widths of \oiii\ in excess of 2500\kms\ while being radio-quiet, and they naturally lie on the extrapolation of the \oiii\ width versus radio relation found in other radio-quiet quasars \citep{Zakamska2014}. Therefore, from kinematics, quasar-driven winds give a more natural explanation to the correlation.

Despite the wide range of radio luminosities observed in AGNs, there exists a correlation between the bolometric luminosity and the median radio luminosity which is characteristic of the radio-quiet population \citep{White2007}. This correlation has been interpreted as evidence that radio emission is due to low-power jets connected to accretion activity \citep{Falcke1995a}. Continuing to explore the jet hypothesis, we turn to the spectral index which is used as a measure of the orientation of jets \citep{Orr1982, Richards2001}. If a jet is viewed perpendicular to its axis, the optically thin synchrotron emission from the lobes and hotspots dominates the observed flux and makes the observed radio spectrum steep. Since the color selection of ERQs is designed to preferentially select obscured objects seen `edge-on' in the classical unification model, we might expect to see steep spectra in jet-dominated ERQs. 

Radio spectra can be further steepened by inverse Compton scattering. At the redshifts of ERQs and on the scales of their host galaxy, the radiation field is dominated by the quasar rather than the commonly considered cosmic microwave background \citep{Krolik1991}. Inverse-Compton losses in particular would affect low-radio-power sources more strongly than high-radio-power sources, which in the compact jet hypothesis might explain why radio-quiet ERQs have steeper spectra ($\alpha\la -1.0$) than radio-loud broad absorption line quasars ($\alpha\simeq -0.7$; \citealt{Becker2000, Richards2001}). We find that the quasar radiation from ERQs is strong enough that inverse Compton scattering can be an important cooling source for the synchrotron-emitting particles \citep{Nims2015}, but its strong dependence on the magnetic field makes it difficult to quantify the effect at this time. The injection spectrum can also vary \citep{Carilli1991}, and in particular in some ultrasteep-spectrum radio sources ($\alpha<-1$) the steepness may be due to a higher density environment into which jets propagate \citep{Klamer2006, Bornancini2010}. In short, the observed radio spectra of jets can be steepened by orientation, steep-spectrum injection, denser environment, rapid synchrotron and inverse-Compton cooling. 

Observations with higher angular resolution could help distinguish compact jets from winds. However, despite dramatically different morphologies and launching mechanisms on scales near the black hole, these two scenarios are surprisingly difficult to disentangle on galaxy scales \citep{Zakamska2014, Harrison2015, Alexandroff2016}. When a strongly collimated jet interacts with the dense interstellar medium of the host galaxy, it loses its directionality and on large scales it appears as an overpressured cocoon which continues to expand and sweep the surrounding medium \citep{Begelman1989}, much like a wind. 

Overall, we are not able to completely rule out compact jets as the origin of the ERQ radio emission. However, we find it unlikely that the optical and infrared selection criteria of ERQs would correlate so strongly with the presence of weak compact steep-spectrum jets, rare in other AGN populations. Furthermore, the compact jet hypothesis does not provide a quantitative explanation of the radio versus kinematics correlation presented in Figure \ref{fig:oIII_radio}. Although it is still possible that the ERQ selection hits the `sweet spot' of weak compact jets, we find that by-product radio emission from winds is a more compelling explanation for the radio emission from ERQs.

\section{Conclusions}
\label{sec:conc}

Extremely red quasars are selected by their red optical-to-mid-infrared colors and high rest equivalent width of \CIV\ \citep{Ross2015, Hamann2017}. They are among the most powerful quasars in the Universe, with bolometric luminosities of $10^{47-48}$\ergs, indicating that ERQs might be accreting at near- or super-Eddington levels.  As many ERQs possess strong outflow signatures traced by ionized gas (\citealt{Zakamska2016}; Perrotta et al. in prep.), ERQs serve as a natural population to explore the connection between radio emission and quasar-driven winds. In this paper, we present radio follow-up observations of 101 ERQs with VLA which we combine with detections in the FIRST survey. We present the following findings: 

\begin{enumerate}
\item $\sim 78$ per cent of ERQs have radio luminosities between $10^{40}-10^{42}$\ergs\ at 6\,GHz, with a median of $1.1\times10^{41}$\ergs. They have steep radio spectra with an average spectral index of $\alpha \sim -1.0$, and many individual objects have even steeper radio spectra. Except for SDSSJ1117+4623, which may be an $\sim 80$\,kpc core+lobe system, no ERQ is resolved with a resolution of $\sim 10$\,kpc.

\item Based on radio morphology, we rule out large-scale jets as the origin of the radio emission in ERQs. We further rule out star formation as the dominant source of radio emission based on unrealistically high required star formation rates and unusually steep spectra of ERQs. Sub-parsec radio coronae and jet cores are also excluded due to their flat expected spectra. 

\item ERQs tightly follow the correlation between the velocity width of \oiii\ and radio luminosity, first established in low-redshift type 2 quasars \citep{Zakamska2014}, suggesting a strong connection between the galactic winds and the radio synchrotron emission. While we cannot completely rule out compact steep-spectrum jets as the origin of ERQs' radio emission, they do not provide a quantitative explanation for this relationship, whereas the observed radio emission can be naturally produced as a by-product of quasar-driven winds propagating into the host galaxy and shocking the surrounding interstellar medium \citep{Jiang2010, Nims2015}. 

\item The wind scenario can explain the observed radio emission if the conversion from the bolometric luminosity to the kinetic power of the wind is $\sim 1$ per cent, consistent with other studies of quasar feedback. In this scenario radio spectra could be steep due to steep-spectrum injection or rapid cooling via synchrotron emission or inverse Compton scattering off a quasar's radiation field.

\item We see evidence that the radio luminosity function of AGNs includes at least three different contributions. All powerful, spatially resolved jet-powered radio sources are on the radio-loud end of the distribution, and the traditional radio-loudness definitions may be helpful in delineating this population. On the low-luminosity end, star formation in the host galaxy can make a dominant contribution. The nature of radio emission with luminosities $10^{39}-10^{42}$\ergs\ remains ambiguous. The discovery of the relationship between radio emission and \oiii\ kinematics in type 2 quasars and ERQs in this range of luminosities suggests that galactic winds in powerful quasars can contribute radio emission in this intermediate regime.

\end{enumerate}

\acknowledgments

The authors are grateful to R. White, S. van Velzen, and U. Rau for technical discussions, to the anonymous referee for constructive suggestions, and to the staff of the National Radio Astronomical Observatory for technical support and assistance with data storage and processing. HCH would also like to acknowledge the helpful discussion with Y.-K. Chiang, A.-L. Sun, and K. Hall. NLZ is grateful to Johns Hopkins University for support via the Catalyst Award and to the Institute for Advanced Study for hospitality during multiple visits. RA is supported by a Postdoctoral Fellowship awarded by the Natural Science and Engineering Research Council of Canada. Support for this work was also provided in part by the National Aeronautics and Space Administration through Chandra Award Number GO6-17100X issued by the Chandra X-ray Observatory Center, which is operated by the Smithsonian Astrophysical Observatory for and on behalf of the National Aeronautics Space Administration under contract NAS8-03060. 


\bibliography{ERQ}{}
\bibliographystyle{aasjournal}

\onecolumngrid
\newpage

\begin{deluxetable*}{cccccc}
\tablecaption{FIRST detections of ERQs. 
\label{table:FIRST-data}
}
\tablehead{\colhead{Source name} & 
\colhead{$z$\tablenotemark{a}} & 
\colhead{$F_{\nu,P}$\tablenotemark{b}} & \colhead{$F_{\nu,I}$\tablenotemark{c}} & \colhead{$L_{radio}$\tablenotemark{d}}  & \colhead{targeting\tablenotemark{e}} }
\startdata
SDSSJ082653.42+054247.3 				    & 2.573     & 1.14  & 0.9	   & 7.10E+41 & \\
SDSSJ083200.20+161500.3 				    & 2.447     & 1.00  & 1.13	   & 7.98E+41 &   \\
SDSSJ091508.45+561316.0 				    & 2.859     & 1.19  & 1.29	   & 1.28E+42 & radio-only\\
SDSSJ095033.51+211729.1 				    & 2.745     & 1.52  & 1.01	   & 9.19E+41 & radio-only\\
SDSSJ095823.14+500018.1 				    & 2.375     & 10.33 & 10.72    & 7.09E+42 &   \\
SDSSJ102447.32$-$013633.8 				& 2.876     & 9.31  & 9.55     & 9.64E+42 & \\
SDSSJ102541.78+245424.2                  & 2.571     & 1.27  & 0.60     & 4.01E+41 & \\
SDSSJ155057.71+080652.1\tablenotemark{f} & 2.514     & 1.28  & 2.36     & 1.77E+42 & \\
SDSSJ164725.72+522948.6 				    & 2.719     & 1.17  & 1.14     & 1.02E+42 & \\
SDSSJ165202.64+172852.3 				    & 2.942	    & 1.64  & 1.26     & 1.34E+42 & \\
\enddata
\tablenotetext{a}{Emission-line redshift from DR12Q.}
\tablenotetext{b}{Peak flux density observed at 1.4\,GHz in\,mJy\pbeam in the FIRST catalog.}
\tablenotetext{c}{Integrated flux density observed at 1.4\,GHz in\,mJy in the FIRST catalog.}
\tablenotetext{d}{Radio luminosity $\nu L_\nu$ at rest-frame 6\,GHz in\ergs.}
\tablenotetext{e}{If a source is marked as radio-only, it is targeted for follow-up SDSS spectroscopy only because of its radio emission detected by FIRST. Radio-only ERQs are excluded from the statistical analysis in the paper.}
\tablenotetext{f}{ERQ-like object.}
\end{deluxetable*}

\begin{deluxetable*}{cccccc}
\tablecaption{Tests of flux density measurement for artificial point sources, as described in Section~\ref{sec:flux-measurement}.  
\label{table:cleanbias}
}
\tablehead{\colhead{$F_{\nu,expected}$\tablenotemark{a}} & 
           \colhead{$F_{\nu,max}$\tablenotemark{b}} & 
           \colhead{$F_{\nu,I, sky}$\tablenotemark{c}} & 
           \colhead{$F_{\nu,P, sky}$\tablenotemark{d}} & 
           \colhead{$F_{\nu,I}$\tablenotemark{e}} & 
           \colhead{$F_{\nu,P}$\tablenotemark{f}}
           }
\startdata
50 	& 58.0 $\pm$5.9 & 61.9$\pm$17.6  & 53.4$\pm$7.5  & 78.9$\pm$12.5  & 55.9$\pm$6.0 \\
100 & 106.9$\pm$6.1 & 107.7$\pm$16.5 & 101.3$\pm$7.2 & 128.6$\pm$11.1 & 103.4$\pm$6.5 \\
200 & 205.6$\pm$6.1 & 204.2$\pm$14.4 & 200.7$\pm$7.0 & 228.1$\pm$10.6 & 201.8$\pm$6.9 \\
\enddata
\tablenotetext{a}{The expected flux density for the artificial point source in\uJy.}
\tablenotetext{b}{Maximum pixel intensity of a source in\uJy\pbeam.}
\tablenotetext{c}{Fitted integrated flux density with consideration of the sky level in\uJy.}
\tablenotetext{d}{Fitted peak flux density with consideration of the sky level in\uJy\pbeam.}
\tablenotetext{e}{Fitted integrated flux density without consideration of the sky level in\uJy.}
\tablenotetext{f}{Fitted peak flux density without consideration of the sky level in\uJy\pbeam.}
\end{deluxetable*}
\onecolumngrid
\newpage

\begin{deluxetable*}{cccccc}
\tablecaption{Spectral Indices, in the order of radio luminosity.
\label{table:spectral-indices}
}
\tablehead{\colhead{Source name}  & 
           \colhead{ERQ type} & 
           \colhead{$L_{radio}$\tablenotemark{a}} & 
           \colhead{$\alpha_{\rm outband}$} & 
           \colhead{$\alpha_{\rm MFS}$} & 
           \colhead{$\alpha_{\rm inband}$} }
\startdata
SDSSJ1117+4623 & main  & 1.9E42 &   $0.17\pm0.10$     &  $0.29\pm0.15$ & $0.37 \pm 0.09$ \\
SDSSJ0209+3122 & main  & 8.1E41 & --\tablenotemark{b} & $-1.86\pm0.39$ & $-1.89 \pm 0.24$ \\
SDSSJ1550+0806 & like  & 7.0E41 & $-1.02\pm0.07$	 & $-1.27\pm0.27$ & $-1.80 \pm 0.72$ \\
SDSSJ0220+0137 & main  & 6.2E41 & $-0.59\pm0.23$	 & $-1.07\pm0.24$ & $-1.12 \pm 0.25$ \\
SDSSJ1145+5742 & like  & 5.2E41 & $-0.97\pm0.13$	 & $-1.15\pm0.17$ & $-0.72 \pm 0.60$ \\
SDSSJ1301+1312 & main  & 4.3E41 & $-1.13\pm0.12$	 & $-0.76\pm0.20$ & $-1.44 \pm 0.27$ \\
SDSSJ1451+2338 & like  & 4.1E41 & $-1.07\pm0.12$	 & $-1.31\pm0.36$ & $-1.31 \pm 0.21$ \\
SDSSJ2215$-$0056\tablenotemark{c} & main   & 1.8E41 & $-1.00\pm0.11$	 & -- & --  \\
Average (80 -- 150\uJy)  & -- &  2.3E41  & $-0.92\pm0.30$ & -- & $-1.08 \pm 0.25$ \\ 
Average ($<$ 80\uJy) & --  &  9.9E40  &   $-1.04\pm0.44$ & -- & $-1.35 \pm 0.34$ \\
\enddata
\tablenotetext{a}{Radio luminosity $\nu L_\nu$ at rest-frame 6\,GHz in\ergs.}
\tablenotetext{b}{Not covered by FIRST.}
\tablenotetext{c}{Detected in Stripe 82.}
\end{deluxetable*}

\onecolumngrid
\newpage

\begin{deluxetable}{ccccc}
\tablecaption{Radio flux densities and luminosities of ERQs.
\label{table:flux}}
\tablecolumns{5}
\tablehead{
\colhead{Source name} & 
\colhead{ ERQ type\tablenotemark{a}} & \colhead{$z$\tablenotemark{b}} & \colhead{$F_\nu$\tablenotemark{c}} & \colhead{$L_{radio}$\tablenotemark{d}}
}
\startdata
SDSSJ000610.67+121501.2 & main & 2.310 & $82   \pm 12$  & 1.45E+41 \\
SDSSJ000746.19+122223.9 & main & 2.429 & $39   \pm 7 $ & 7.73E+40 \\
SDSSJ001120.22+260109.2 & like & 2.290 & $64   \pm 8 $ & 1.11E+41 \\
SDSSJ002400.25+245031.9 & main & 2.809 & $40   \pm 8 $ & 1.10E+41 \\
SDSSJ003747.98+170504.0 & like & 3.664 & $47   \pm 8 $ & 2.27E+41 \\
SDSSJ004713.21+264024.7 & main & 2.564 & $49   \pm 8 $ & 1.09E+41 \\
SDSSJ005044.95$-$021217.6 & main & 2.254 & $     < 29  $ & $<$4.88E+40 \\
SDSSJ005233.24$-$055653.5 & main & 2.354 & $134  \pm 8 $ & 2.47E+41 \\
SDSSJ014111.13$-$031852.5 & main & 2.559 & $45   \pm 8 $ & 9.93E+40 \\
SDSSJ015222.58+323152.7 & main & 2.786 & $     < 27  $ & $<$7.20E+40 \\
SDSSJ020932.15+312202.7 & main & 2.376 & $429  \pm 8 $ & 8.06E+41 \\
SDSSJ022052.11+013711.1 & main & 3.138 & $199  \pm 7 $ & 6.91E+41 \\
SDSSJ080425.75+470159.0 & main & 2.772 & $     <  47 $ & $<$1.25E+41 \\
SDSSJ080547.66+454159.0 & main & 2.326 & $62   \pm 8 $ & 1.11E+41 \\
SDSSJ082536.31+200040.3 & main & 2.094 & $     < 30  $ & $<$4.32E+40 \\
SDSSJ082508.77+355822.9 & like & 2.894 & $63   \pm 8 $ & 1.83E+41 \\
SDSSJ082649.30+163945.2 & main & 2.316 & $     < 29  $ & $<$5.24E+40 \\
SDSSJ083448.48+015921.1 & main & 2.594 & $83   \pm 8 $ & 1.90E+41 \\
SDSSJ084447.66+462338.7 & main & 2.217 & $103  \pm 8 $ & 1.66E+41 \\
SDSSJ084424.50+545234.2 & like & 2.702 & $34   \pm 8 $ & 8.50E+40 \\
SDSSJ085451.11+173009.1 & main & 2.614 & $     < 32  $ & $<$7.48E+40 \\
SDSSJ090306.18+234909.8 & main & 2.263 & $     < 32  $ & $<$5.47E+40 \\
SDSSJ091303.90+234435.2 & main & 2.420 & $50   \pm 9 $ & 9.72E+40 \\
SDSSJ092049.59+282200.9 & main & 2.296 & $     < 32 $  & $<$5.60E+40 \\
SDSSJ093226.93+461442.8 & main & 2.314 & $53   \pm 9 $ & 9.31E+40 \\
SDSSJ093506.96$-$024137.7 & main & 2.167 & $     < 36 $ & $<$5.56E+40 \\
SDSSJ093638.41+101930.3 & main & 2.453 & $     < 53 $  & $<$1.07E+41 \\
SDSSJ093926.67+591311.1 & like & 3.744 & $     < 30 $  & $<$1.52E+41 \\
SDSSJ101324.53+342702.6 & main & 2.482 & $52   \pm 10$ & 1.08E+41 \\
SDSSJ101326.23+611219.9 & like & 3.703 & $101  \pm 7 $ & 4.98E+41 \\
SDSSJ101533.65+631752.6 & main & 2.226 & $     < 32 $  & $<$5.12E+40 \\
SDSSJ102130.74+214438.4 & main & 2.194 & $35   \pm 8 $ & 5.52E+40 \\
SDSSJ102353.44+580004.9 & main & 2.597 & $23   \pm 6 $ & 5.36E+40 \\
SDSSJ103146.53+290324.1 & main & 2.293 & $61   \pm 7 $ & 1.06E+41 \\
SDSSJ103456.95+143012.5 & main & 2.961 & $65   \pm 8 $ & 1.98E+41 \\
SDSSJ104611.50+024351.6 & main & 2.773 & $88   \pm 7 $ & 2.32E+41 \\
SDSSJ104718.35+484433.8 & main & 2.275 & $     < 28  $ & $<$4.73E+40 \\
SDSSJ104754.58+621300.5 & main & 2.536 & $48   \pm 7 $ & 1.05E+41 \\
SDSSJ110202.68$-$000752.7 & main & 2.626 & $     < 32  $ & $<$7.40E+40 \\
SDSSJ111017.13+193012.5 & main & 2.505 & $42   \pm 7 $ & 8.85E+40 \\
SDSSJ111346.10+185451.9 & main & 2.516 & $     < 27  $ & $<$5.68E+40 \\
SDSSJ111355.72+451452.6 & main & 2.188 & $109  \pm 7 $ & 1.70E+41 \\
SDSSJ111516.33+194950.4 & main & 2.792 & $     < 27  $ & $<$7.22E+40 \\
SDSSJ111729.56+462331.2 & main & 2.132 & $1266 \pm 7 $ & 1.87E+42 \\
SDSSJ113349.71+634740.0 & main & 2.202 & $122  \pm 7 $ & 1.94E+41 \\
SDSSJ113834.68+473250.0 & main & 2.310 & $47   \pm 7 $ & 8.27E+40 \\
SDSSJ114508.00+574258.6 & like & 2.790 & $145  \pm 7 $ & 3.88E+41 \\
SDSSJ114855.47+572159.8 & like & 3.054 & $     < 28  $ & $<$9.02E+40 \\
SDSSJ121704.70+023417.1 & main & 2.416 & $30   \pm 7 $ & 5.92E+40 \\
SDSSJ122000.68+064045.3 & main & 2.796 & $     < 28  $ & $<$7.43E+40 \\
SDSSJ123241.73+091209.3 & main & 2.381 & $37   \pm 7 $ & 7.05E+40 \\
SDSSJ124106.97+295220.8 & main & 2.794 & $34   \pm 7 $ & 9.12E+40 \\
SDSSJ124738.40+501517.7 & main & 2.386 & $42   \pm 7 $ & 8.06E+40 \\
SDSSJ125019.46+630638.6 & main & 2.402 & $28   \pm 7 $ & 5.30E+40 \\
SDSSJ125449.50+210448.4 & main & 3.118 & $33   \pm 7 $ & 1.14E+41 \\
SDSSJ125811.25+212359.6 & main & 2.614 & $36   \pm 8 $ & 8.34E+40 \\
SDSSJ130114.46+131207.4 & main & 2.787 & $147  \pm 7 $ & 3.92E+41 \\
SDSSJ130421.12+083752.2 & main & 2.946 & $162  \pm 30$   & 4.88E+41 \\
SDSSJ130654.76+132704.8 & main & 2.499 & $60   \pm 7 $ & 1.26E+41 \\
SDSSJ130630.66+584734.7 & main & 2.297 & $28   \pm 8 $ & 4.86E+40 \\
SDSSJ130936.14+560111.3 & main & 2.569 & $50   \pm 16$  & 1.12E+41 \\
SDSSJ131047.78+322518.3 & main & 3.017 & $55   \pm 29$   & 1.76E+41 \\
SDSSJ131722.85+322207.5 & main & 2.400 & $33   \pm 8 $ & 6.39E+40 \\
SDSSJ131833.76+261746.9 & main & 2.272 & $21   \pm 7 $ & 3.55E+40 \\
SDSSJ134026.99+083427.2 & main & 2.483 & $50   \pm 8 $ & 1.04E+41 \\
SDSSJ134001.90+322155.9 & main & 2.405 & $     < 30  $ & $<$5.85E+40 \\
SDSSJ134450.51+140139.2 & main & 2.745 & $     < 32  $ & $<$8.40E+40 \\
SDSSJ134417.34+445459.4 & main & 3.036 & $     < 90  $ & $<$2.89E+41 \\
SDSSJ134535.66+600028.4 & main & 2.937 & $     < 26  $ & $<$7.86E+40 \\
SDSSJ135557.60+144733.1 & main & 2.704 & $     < 29  $ & $<$7.25E+40 \\
SDSSJ135608.32+073017.2 & main & 2.269 & $     < 34  $ & $<$5.74E+40 \\
SDSSJ140506.80+543227.3 & main & 3.213 & $68   \pm 8 $ & 2.48E+41 \\
SDSSJ141350.76+214307.7 & main & 2.440 & $     < 27  $ & $<$5.47E+40 \\
SDSSJ143159.76+173032.6 & main & 2.377 & $     < 39  $ & $<$7.36E+40 \\
SDSSJ144742.66+565209.2 & like & 2.789 & $97   \pm 7 $ & 2.58E+41 \\
SDSSJ145113.61+013234.1 & like & 2.773 & $58   \pm 8 $ & 1.53E+41 \\
SDSSJ145148.01+233845.4 & like & 2.621 & $167  \pm 7 $ & 3.90E+41 \\
SDSSJ145354.70+190343.9 & main & 2.349 & $87   \pm 10$  & 1.59E+41 \\
SDSSJ145623.35+214516.2 & main & 2.476 & $80   \pm 12$  & 1.65E+41 \\
SDSSJ150117.07+231730.9 & main & 3.025 & $     < 39  $  & $<$1.26E+41 \\
SDSSJ152941.01+464517.6 & main & 2.420 & $     < 43  $  & $<$8.38E+40 \\
SDSSJ153107.14+105825.8 & main & 2.781 & $     < 32  $ & $<$8.51E+40 \\
SDSSJ153108.10+213725.1 & main & 2.569 & $     < 34  $ & $<$7.60E+40 \\
SDSSJ153446.26+515933.8 & main & 2.265 & $80   \pm 9 $ & 1.36E+41 \\
SDSSJ154243.87+102001.5 & main & 3.215 & $47   \pm 8 $ & 1.70E+41 \\
SDSSJ154523.97+165054.2 & like & 2.954 & $     < 32  $ & $<$9.85E+40 \\
SDSSJ154743.78+615431.1 & main & 2.868 & $57   \pm 10$  & 1.61E+41 \\
SDSSJ154831.92+311951.4 & main & 2.736 & $40   \pm 9 $ & 1.02E+41 \\
SDSSJ155057.71+080652.1 & like & 2.514 & $266  \pm 8 $ & 5.67E+41 \\
SDSSJ160431.55+563354.2 & main & 2.484 & $62   \pm 8 $ & 1.29E+41 \\
SDSSJ162920.36+495705.3 & like & 2.759 & $93   \pm 8 $ & 2.44E+41 \\
SDSSJ170558.64+273624.7 & main & 2.448 & $34   \pm 7 $ & 6.88E+40 \\
SDSSJ171420.38+414815.7 & main & 2.342 & $42   \pm 8 $ & 7.64E+40 \\
SDSSJ220337.79+121955.3 & main & 2.623 & $102  \pm 29$   & 2.38E+41 \\
SDSSJ221524.00$-$005643.8 & main & 2.509 & $98   \pm 7 $ & 2.09E+41 \\
SDSSJ222421.63+174041.2 & main & 2.165 & $     < 28  $ & $<$4.31E+40 \\
SDSSJ223754.52+065026.6 & main & 2.609 & $     < 28  $ & $<$6.44E+40 \\
SDSSJ225438.30+232714.5 & main & 3.087 & $     < 28  $ & $<$9.47E+40 \\
SDSSJ232326.17$-$010033.1 & main & 2.356 & $     < 34  $ & $<$6.31E+40 \\
SDSSJ232828.47+044346.8 & main & 2.564 & $     < 48  $  & $<$1.05E+41 \\
SDSSJ233636.99+065231.0 & main & 2.776 & $54   \pm 7 $ & 1.44E+41 \\
\enddata
\tablenotetext{a}{`Main' sources are refer to ERQs in this paper, and to `core' ERQs in \cite{Hamann2017}. `Like' sources are ERQ-like objects in both \cite{Hamann2017} and this paper.}
\tablenotetext{b}{Emission-line redshift from DR12Q.}
\tablenotetext{c}{Flux density observed at 6.2\,GHz in\uJy. Flux density error is the root-mean-square noise of each field based on the test in Section~\ref{sec:flux-measurement}.}
\tablenotetext{d}{Radio luminosity $\nu L_\nu$ at rest-frame 6\,GHz in\ergs.}
\end{deluxetable}

\end{document}